\documentclass{jfm}
\usepackage{graphicx}
\usepackage{epstopdf, epsfig}

\usepackage{amsmath}
\usepackage{placeins}
\usepackage{xcolor}

\newcommand\Capi{\mbox{\textit{Ca}}} 
\newcommand\Bon{\mbox{\textit{Bo}}}  

\shorttitle{Bubble dynamics in a geometrically-perturbed Hele-Shaw channel} 
\shortauthor{A. Gaillard et al.} 

\title{The life and fate of a bubble in a geometrically-perturbed Hele-Shaw channel}

\author
{
	Antoine Gaillard\aff{1},
	Jack S. Keeler\aff{2},
	Gr\'egoire Le Lay\aff{3},
	Gr\'egoire Lemoult\aff{4},
	Alice B. Thompson\aff{5},
	Andrew L. Hazel\aff{5},
	\and 
	Anne Juel\aff{1}
	\corresp{\email{anne.juel@manchester.ac.uk}}
}

\affiliation
{
	\aff{1}
	Manchester Centre for Nonlinear Dynamics and Department of Physics and Astronomy, University of Manchester, Oxford Road, Manchester M13 9PL, UK
	\aff{2}
	Mathematics Institute, University of Warwick, Coventry CV4 7AL, UK
	\aff{3}
	Department of Physics, \'Ecole Normale Sup\'erieure, 24 rue Lhomond, 75005 Paris, France
	\aff{4}
	CNRS, UMR 6294, Laboratoire Onde et Milieux Complexes (LOMC)53, Normandie Universit\'e, UniHavre, rue de Prony, Le Havre Cedex76058, France
	\aff{5}
	Department of Mathematics and Manchester Centre for Nonlinear Dynamics and University of Manchester, Oxford Road, Manchester M13 9PL, UK
}

\begin{document}
\maketitle

\begin{abstract}

Motivated by the desire to understand complex transient behaviour in fluid flows, we study the dynamics of an air bubble driven by the steady motion of a suspending viscous fluid within a Hele-Shaw channel with a centred depth perturbation. Using both experiments and numerical simulations of a depth-averaged model, we investigate the evolution of an initially centred bubble of prescribed volume as a function of flow rate and initial shape. The experiments exhibit a rich variety of organised transient dynamics, involving bubble break up as well as aggregation and coalescence of interacting neighbouring bubbles. The long-term outcome is either a single bubble or multiple separating bubbles, positioned along the channel in order of increasing velocity. Up to moderate flow rates, the life and fate of the bubble are reproducible and can be categorised by a small number of characteristic behaviours that occur in simply-connected regions of the parameter plane. Increasing the flow rate leads to less reproducible time evolutions with increasing sensitivity to initial conditions and perturbations in the channel. Time-dependent numerical simulations that allow for break up and coalescence are found to reproduce most of the dynamical behaviour observed experimentally including enhanced sensitivity at high flow rate. An unusual feature of this system is that the set of steady and periodic solutions can change during temporal evolution because both the number of bubbles and their size distribution evolve due to break up and coalescence events. Calculation of stable and unstable solutions in the single- and two-bubble cases reveals that the transient dynamics are orchestrated by weakly-unstable solutions of the system termed edge states that can appear and disappear as the number of bubbles changes.

\end{abstract}


\section{Introduction}

Fundamental studies of complex fluid mechanical phenomena usually concentrate on the steady or periodic flows that are observed after some time. Experimental observation of steady or periodic flows is used to infer that they are stable solutions of the underlying governing equations. Unstable solutions of the governing equations, on the other hand, are often ignored being thought to be unimportant and unobservable in experiments. In fact, unstable solutions can be glimpsed fleetingly in experiments as they influence the transient evolution of a flow, which is often as important as the final state.

In this paper, we conduct detailed experiments to characterise the transient evolution and ultimate disposition of a simple fluid mechanical system that nonetheless exhibits complex nonlinear behaviour: a finite air bubble propagating within a liquid-filled Hele-Shaw channel containing a relatively small depth perturbation; see figure \ref{fig:setup}. 
The stability of finite steady bubbles in unperturbed channels has previously been addressed both experimentally \citep{homsy} and numerically \citep{tanveer1987stability}. However, interfacial flows in narrow gaps can also exhibit considerable disorder \citep{saffman1958penetration,Couder2000,Christian}, but they are rarely investigated from a dynamical systems perspective. Although our primary interest is to use the system as a model in which to study complex transient dynamics, the flow of bubbles through channels also has a number of industrial, chemical and biological applications, particularly in microfluidic geometries \citep{Anna2016}.

The transient behaviour of a flow is particularly significant when the flow is disordered with no apparent stable steady or periodic states. The connection between disorder in fluid mechanics and the nascent theory of dynamical systems was initiated via low-dimensional mathematical models of turbulence \citep{Hopf1948} and atmospheric convection \citep{Lorenz1963}. Subsequently, \citet{Ruelle} identified turbulence with deterministic chaos. Characterised by long-term unpredictability due to sensitivity to initial conditions in systems of three or more degrees of freedom, chaos was shown to emerge after only a small number of bifurcations. The first observation of chaos in a fluid system was made by \citet{Swinney} in Taylor--Couette flow, where geometric confinement reduces the spatial complexity of the flow \citep{Mullin}, so that its temporal dynamics can be captured by models of reduced dimensionality tracking a small number of dominant eigenmodes that control the dynamics \citep{Manneville}. The relaxation of spatial constraints by increasing the aspect ratio of such closed flows can lead to transitions to spatio-temporal chaos \citep{Ahlers, Gollub1995}. However, theoretical insight into the dynamics of these systems tends to be reduced because of the increased number of significant eigenmodes. In recent years, interest in dynamical systems ideas in fluid dynamics has been rekindled through their considerable success in elucidating the subcritical transition to turbulence in shear flows \citep{kerswell2005recent,schneider2007turbulence,eckhardt2008dynamical,duguet2008}. Rather than focusing on the eigenmodes of these weakly-confined, open flows, the approach has been to reduce the dynamics to a small number of invariant objects in phase space. 

It is now widely accepted that stable and unstable invariant objects play a key role in organising the dynamics of transition to turbulence \citep{kawahara2012,cvitanovic_2013}. In this context, an invariant object is a subset of the possible states of the system that does not change under time evolution. For example, a steady solution is invariant because the system does not move away from it unless perturbed. A periodic solution is also invariant because the system evolves through a limited set of states over one period and then the cycle repeats. More complex invariant objects include the stable and unstable manifolds associated with steady and periodic solutions and chaotic attractors. The fact that the system cannot leave an invariant state means that they divide the possible states of the system into regions with different dynamical outcomes; in other words, trajectories of the system cannot cross an invariant object. For the turbulent transition problem, the most important invariant is the so-called `edge' of turbulence dividing regions that eventually return to laminar flow from those that progress to a fully turbulent state.

A simple conceptual picture of an edge state is to consider the two-dimensional steady stagnation point flow of a fluid. The stagnation point itself is an unstable fixed point of the flow because fluid particles will move away if displaced from the stagnation point. As well as the stagnation point, the important invariant objects are the stable and unstable manifolds, which are simply the dividing streamlines of the flow. These manifolds divide the flow into four different regions and a particle that starts in one region cannot move into another. Hence a complete characterisation of the ultimate fate of a fluid particle follows from knowing its position relative to the dividing streamlines. In other words, the manifolds are the edge states of this simple system.

A difficulty in applying these ideas in fluid dynamics is that the invariant objects of interest can be extremely high dimensional and cannot be easily determined. Nevertheless, recent advances in both computer power and efficient algorithms \citep{Tuckerman2019,farano_etal_2019}
have allowed ever increasing access to such objects in studies of turbulence \citep{budanur2017}, convection \citep{Sanchez_2019} and droplet break up \citep{gallino2018edge}.

In our chosen system, the bubble exhibits a variety of complex behaviours including symmetry breaking, bistability, steady and periodic states as well as non-trivial transient evolution \citep{franco2018bubble}. The Reynolds number remains small and the nonlinearities arise only through the presence of the air-liquid interface. Hence, direct observation of the interface can be used to distinguish different possible states of the system. We develop a new experimental protocol to generate a variety of different initial bubble configurations and then investigate the bubble's subsequent behaviour under imposition of flow. We find a rich variety of transient evolutions that can, nonetheless, be rationalised and classified because they occur in distinct regions of the parameter space. 

In addition the system is attractive from a theoretical point of view because its behaviour can be captured both qualitatively and quantitatively by a depth-averaged set of equations \citep{franco2016sensitivity}, provided that the aspect ratio of the channel's cross-section is sufficiently large.
\citet{keeler2019influence} presented a purely theoretical study of the depth-averaged model in which unstable periodic orbits were found to be edge states marking the boundary between steady propagation along either the centreline (symmetric) or biased towards one side of the channel (asymmetric). In the present paper, we synthesise experimental data and numerical solutions of the depth-averaged equations to examine the influence of a wide variety of stable and unstable invariant objects on the behaviour of the bubble.


A distinct feature of the system is the propensity of the bubble to break up into two or more smaller bubbles that may or may not recombine in the subsequent dynamics. The transition from single to multiple bubbles allows the evolution of the invariant-object structure for fixed parameter values, which is rather unusual. The resulting complex dynamics are nonetheless orchestrated by a variety of edge states that are invariant solutions of the depth-averaged system.

The paper is organised as follows. The first three experimental sections present the experimental set-up and protocol (\S \ref{sec:Material and methods}), the method used to set the initial bubble shape (\S \ref{sec:Initial bubble shape}) and the experimental results (\S \ref{sec:observations}). This is followed by a presentation of the depth-averaged model and numerical methods (\S \ref{sec:Depth-averaged model for bubble propagation}) and by an interpretation of the experimental results in terms of dynamical system arguments (\S\S \ref{sec:Transient evolution}--\ref{sec:Interaction between two bubbles of comparable sizes}).
We draw our conclusions in \S \ref{sec:conclusion}.


\section{Experimental methods}
\label{sec:Material and methods}

\subsection{Experimental setup}
\label{sec:Experimental setup}

\begin{figure}
	\centerline{\includegraphics[scale=1]{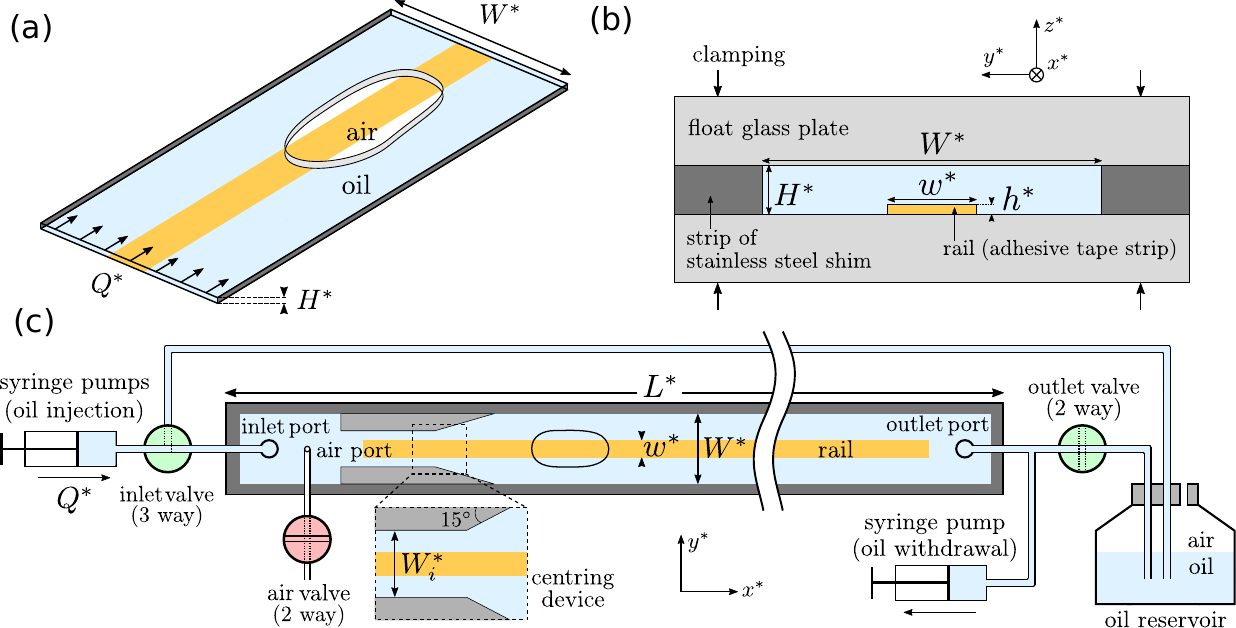}}
	\caption{(a) Schematic diagram of a bubble propagating in the experimental channel. (b) End view of the channel with a centred, rectangular, depth perturbation (in the absence of a bubble). (c) Schematic diagram of the experimental setup.}
	\label{fig:setup}
\end{figure}

A schematic diagram of the experimental setup is shown in figure \ref{fig:setup}. The Hele-Shaw channel used in the experiments consisted of two parallel float glass plates of dimensions 170~cm $\times$ 10~cm $\times$ 2~cm separated by two strips of stainless steel shim. The shims were sprayed with a thin layer of adhesive (3M spray mount\texttrademark) and bonded to the bottom glass plate with a distance $W^*$ between them, which was imposed using gauge blocks. The channel height was $H^* = 1.00 \pm 0.01$~mm and its width was $W^* = 40.0$~mm with variations up to $\pm 0.1$~mm over the length of the channel  $L^* = 170$~cm, yielding an aspect ratio $\alpha = W^*/H^* = 40$. A thin depth perturbation of rectangular cross-section, width $w^* = 10.0 \pm 0.1$~mm and thickness $h^* = 24 \pm 1$~$\mu$m, was positioned on the bottom glass plate, symmetrically about the channel centreline; see figure \ref{fig:setup}(b). This depth perturbation, henceforth referred to as a rail, was made from a translucent adhesive tape strip bonded to the bottom glass plate and its surface gently smoothed with sandpaper of grit size P1500. The strip width was larger than $w^*$ and the excess width on each side was trimmed using a sharp blade placed against a 15~mm gauge block sliding along the fixed steel shims. The top glass plate was then placed onto the shims and the channel was sealed with clamps and levelled horizontally to within 0.03$^{\circ}$.

The channel was filled with silicone oil (Basildon Chemicals Ltd) of dynamic viscosity $\mu = 0.019$~Pa.s, density $\rho = 951$~kg/m$^3$ and surface tension $\sigma = 21$~mN/m at the laboratory temperature of $21\pm 1$~$^{\circ}$C. Three cylindrical brass pieces embedded into the top glass plate provided oil inlet and outlet ports as well as an air injection port (see figure \ref{fig:setup}(c)), and all connections were made with rigid plastic tubes. The output from three syringe pumps in parallel (Legato 200 series) was fed to a three-way solenoid valve connected to the channel inlet and to an oil reservoir at atmospheric pressure, which was also linked to the outlet of the channel via a two-way solenoid valve. Hence, depending on the setting of the three-way inlet valve, oil could be injected into the channel at a constant flow rate $Q^*$ (open outlet valve) or withdrawn from the reservoir to refill the syringes.

The air injection port was connected to ambient air through a two-way solenoid valve. An air bubble was injected into the channel through the air port by closing the outlet valve, opening the air valve and slowly withdrawing a set volume of oil from the channel via a syringe pump connected to the outlet. Once the bubble was formed, the air valve was closed, the outlet valve was opened, and a small amount of oil was slowly injected into the channel to nudge the bubble forward. This prevented the compression of the bubble into the air port when flow was initiated. A bubble centring device was positioned downstream of the bubble injection port, where the width of the channel was reduced to $W^*_i = 17$~mm over a length of 110~mm which expanded linearly over 40~mm into the channel of width $W^* = 40$~mm, thus featuring an expansion ratio of $W^*/W_i^* = 2.4$. The rail started within this device.

A centred bubble of controlled shape was obtained in the channel using a protocol described in \S \ref{sec:Initial bubble shape}. This bubble was then set in motion from rest by imposing a constant oil flow rate $Q^*$. The propagating bubble was filmed in top-view using a CMOS camera mounted on a motorised translation stage, which was placed above the channel and spanned its entire length $L^*$. An empirical relation between the bubble speed and oil flow rate was used to set a constant translation speed for the camera so that the propagating bubble remained within the field of view of the camera for the duration of the experiment. The channel was uniformly back-lit with a custom made LED light box, so that the oil-air interface appeared as a black line through light refraction. The syringe pumps, valves, translation stage and camera were computer-controlled via a LabVIEW code. 

We measured the projected area $A^*$ of the bubble and its centroid position from the bubble contour, which was identified by an edge detection algorithm in terms of $x^*$ and $y^*$ coordinates spanning the length and width of the channel, respectively (see figure \ref{fig:setup}). We will examine the effects of flow on $A^*$ in \S \ref{sec:Effect of the flow on the projected area of the bubble}. We also determined the bubble velocity $U_b^*$ along the $x^*$ direction from the camera speed and the centroid position on successive frames. Henceforth, we will refer to the bubble size in terms of an equivalent radius $r^* = \sqrt{A^*/\upi}$ based on the projected area of the bubble. We only considered bubbles larger than the rail, i.e. $2r^* > w^*$, for which the bubble confinement ratio $r^*/H^*$ was therefore larger than $5$. We denote as $y_c^*$ the distance from the bubble centroid to the centreline of the channel. We choose the channel half-width $W^*/2$ and the average oil velocity in a channel without a rail $U_0^* = Q^*/(W^* H^*)$ as the characteristic length and velocity scales in the $x^*$-$y^*$ plane, respectively. The non-dimensional bubble size, centroid offset and velocity are therefore $r = 2 r^*/W^*$, $y_c = 2y^*_c/W^*$ and $U_b = U_b^*/U_0^*$, respectively. We define a non-dimensional flow rate $Q = \mu U_0^* / \sigma$ which is a capillary number based on the mean oil velocity (in the channel without a rail) and which ranges between $0.005$ and $0.19$ in the experiment. The capillary number based on the bubble velocity, which measures the ratio of viscous to capillary forces, is $\Capi = \mu U_b^* / \sigma$ with values in the range $0.008\le \Capi \le 0.48$. The ratio of inertial to viscous forces yields a reduced Reynolds number $\Rey = \rho U_0^* H^{*2} / \mu W^*$ which takes values less than $0.3$ in our experiments, thus indicating negligible inertia. The Bond number $\Bon = \rho g H^{*2} / 4 \sigma=0.11$, where $g$ is the gravitational acceleration, indicates that bubble buoyancy is also negligible. 



\subsection{Effect of flow on the projected area of the bubble}
\label{sec:Effect of the flow on the projected area of the bubble}

The projected area of a propagating bubble is affected by two competing effects: films of oil separating the bubble from the top and bottom glass plates tend to increase the projected area of the bubble, while the pressure increase associated with the imposition of flow in the channel reduces the projected area through compression of the bubble. 

Films of oil on the top and bottom glass plates encapsulate a newly created bubble because silicone oil wets the channel walls. When the bubble is at rest ($Q^*=0$), these films tend to be very thin, but they drain so slowly that they are always present on the time scale of the experiment. The thickness of these films increases with the capillary number for a propagating bubble \citep{bretherton1961motion}, so that for a fixed bubble volume, the projected area of the bubble increases accordingly. 

We estimated the effect of wetting films in our experiments by measuring their thickness for a bubble propagating in a channel where the rail was absent; see Appendix \ref{sec:Wetting films}. We found that for the maximum value of $\Capi=0.48$ investigated in this paper, the film thickness averaged over the bubble area is $\langle h_f^* \rangle / H^* \approx 0.2$ (i.e. $\langle h_f^* \rangle \approx 200$~$\mu$m). This means that up to $40$~\% of the channel depth is filled with oil, and that the bottom film could be up to $8$ times thicker than the rail ($h^* = 24 \pm 1$~$\mu$m). We expect that the increasing film thickness will lead to a reduced influence of the rail on the bubble propagation as the flow rate increases.

\begin{figure}
	\centerline{\includegraphics[scale=1]{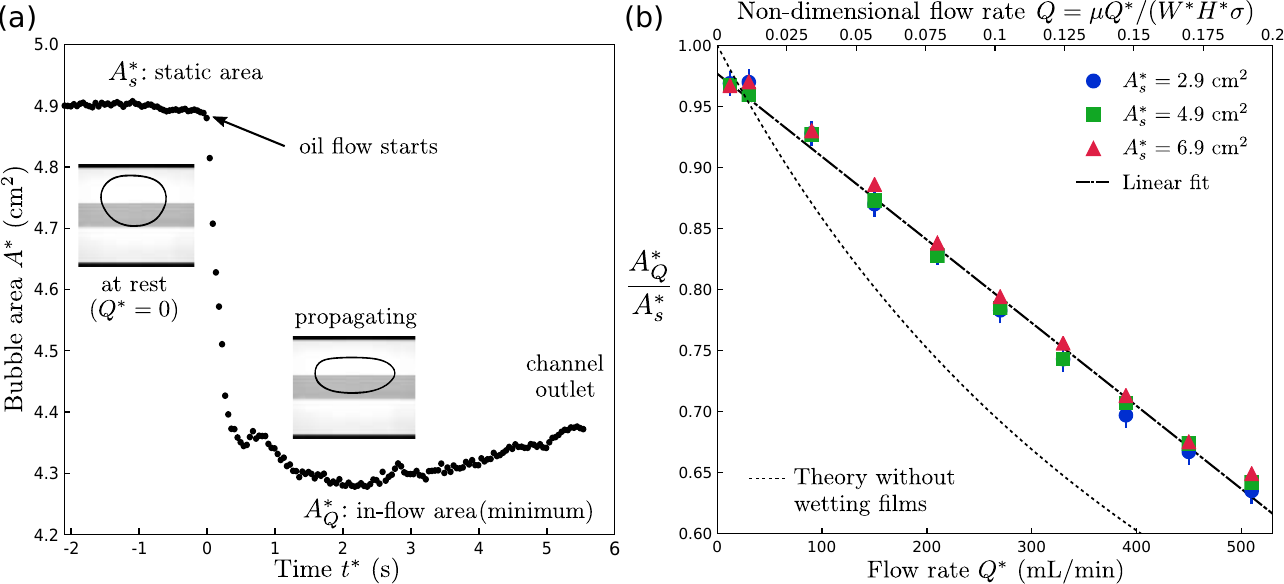}}
	\caption{Bubble compression in flow. (a) Variation of the projected area $A^*$ of a bubble initially at rest ($Q^*=0$), asymmetric about the channel centreline, when a flow rate $Q^* = 150$~mL/min is imposed at $t^*=0$ as indicated by the arrow. The bubble area rapidly decreases from its static value $A_s^* = 4.9$~cm$^2$ to a minimum value $A_Q^*=4.3$~cm$^2$ (13~\% reduction) due to the sudden pressure increase in the channel, and then slowly increases by about 2~\% as it propagates towards the channel outlet where pressure is lower. Inset snapshots indicate the shape of the bubble before and after flow imposition and the grey band is the rail. (b) Compression ratio $A_Q^*/A_s^*$ as a function of the dimensional flow rate $Q^*$ and non-dimensional flow rate $Q$ for three static bubble sizes. The data is accurately captured by the linear relation $A_Q^*/A_s^* = 0.98 - K^* Q^*$ where $K^* = 4.1 \, 10^{4}$~s/m$^3$ (dash-dotted line). The dotted line corresponds to the prediction of the compression ratio when assuming ideal gas behaviour and neglecting wetting films (see main text).}
	\label{fig:compressibility}
\end{figure}

Upon flow initiation, we systematically observed a net reduction of the bubble projected area despite thickening wetting films, which suggests that compression of the bubble due to the imposed flow plays a dominant role. This is illustrated in figure \ref{fig:compressibility}(a) in which a bubble initially at rest ($Q^*=0$) with a static projected area $A_s^* = 4.9$~cm$^2$ compresses rapidly when a flow rate $Q^* = 150$~mL/min is imposed to reach a minimum area of $A_Q^*=4.3$~cm$^2$ (13~\% reduction). This rapid decrease in bubble area is followed by a slow increase of approximately 2~\% as the bubble propagates towards the channel outlet. We chose to propagate strongly asymmetric bubbles to measure compression effects because they persist over the entire range of flow rates investigated, whereas symmetric bubbles are prone to break up when propagating from rest on the rail; see \S \ref{sec:observations}. Such a  bubble was prepared by initially propagating it at a low flow rate $Q^*=10$~mL/min for which it systematically migrated sideways towards one of the deeper sides of the channel before interrupting the flow; see \S \ref{sec:Bifurcation diagram} for a summary of the steady states of the system. 

Figure \ref{fig:compressibility}(b) shows the compression ratio $A_Q^*/A_s^*$ for different flow rates $Q^*$ and three representative static bubble sizes. $A_Q^*/A_s^*$ decreases monotonically and approximately linearly with increasing flow rate and does not depend significantly on the bubble size. A least-square linear fit $A_Q^*/A_s^* = K_0 - K^* Q^*$ yields $K_0 = 0.98$ and $K^* = 4.1 \times 10^{4}$~s/m$^3$. The intercept $A_Q^*/A_s^* =0.98$ for $Q^*=0$ suggests a steep increase towards $A_Q^*/A_s^* =1.00$ for vanishing flow rates, which was also reported by \citet{franco2017bubble}.

The compression ratio $A_Q^*/A_s^*$ can be estimated by considering that the fluid pressure acting on the bubble increases rapidly from atmospheric pressure $P_A^*$ to $P_A^* + G^* l^* + G_t^* l_t^*$ when the flow is imposed. $G^* = 12 \mu Q^* / (W^* H^{*3})$ is the pressure gradient in the channel, $l^*$ is the distance initially separating the bubble from the channel outlet, typically $110$~cm in our experiments and $G_t^* = 8 \mu Q^* / (\upi r_t^{*4})$ is the pressure gradient in the tube of radius $r_t^* = 2$~mm and length $l_t^* = 120$~cm connecting the channel outlet to the reservoir at atmospheric pressure; see figure \ref{fig:setup}. 
Assuming ideal gas behaviour and neglecting wetting films for simplicity, we get a compression ratio $A_Q^* / A_s^* = (1+ (G^*l^* +G_t^*l_t^*)/P^*_A)^{-1}$. This expression is plotted as a function of flow rate in figure \ref{fig:compressibility}(b). The discrepancy with the experimental data is due to the increase in projected area resulting from the increase in wetting film thickness when the bubble propagates, which partly compensates the reduction in bubble volume due to compression.

The bubble does not recover its initial projected area $A_s^*$ when approaching the channel outlet partly because the pressure is still larger than $P_A^*$. However, the remaining pressure head $G_t^* l_t^*$ is too small to account for the modest $2$~\% increase observed in figure \ref{fig:compressibility}(a). The missing area is consistent with air diffusion across the oil-air interface during bubble propagation. Experiments for $A_s^*=3.9$~cm$^2$, where the flow rate was interrupted when the bubble was close to the channel outlet, showed that the projected area of the bubble at rest after propagation was less than the initial value $A_s^*$ by $0.35$~\% to $11$~\% for flow rates ranging from $10$~mL/min to $477$~mL/min. This suggests that the solubility of air in oil increases with pressure, as previously reported by \citet{franco2017bubble}. Hence, in our bubble propagation experiments, air diffusion into the silicone oil helps to retain an approximately constant bubble area after initial compression. The maximum increase in bubble area was $7$~\% for the highest flow rates, which was small enough to avoid a measurable change in bubble propagation; see also \S \ref{sec:Bifurcation diagram} and Appendix \ref{sec:Influence of the bubble size} for a discussion of the influence of the bubble size.

In order to account for these effects, we parametrized our bubble propagation experiments according to the in-flow equivalent radius of the bubble $r_Q = (2 \sqrt{A_Q^*/ \upi}) / W^*$ and we used the relation presented in figure \ref{fig:compressibility}(b) to obtain bubbles with the required value of projected area $A_Q^*$. Note that compression ratios similar to figure \ref{fig:compressibility}(b) were found for the initially symmetric bubbles prepared with the protocol described in \S \ref{sec:Initial bubble shape}.


\section{Relaxation of a bubble at $Q=0$: selection of the initial bubble shape}
\label{sec:Initial bubble shape}

The relaxation of a centred, elongated bubble in the absence of imposed flow provides a non-invasive method to vary the shape of a bubble while conserving its volume. We selected specific shapes reproducibly as initial conditions for the propagation experiments described in \S \ref{sec:observations} by controlling the time for which the bubble relaxes from its original shape before imposing the flow. 

The elongated bubble was prepared by flowing a newly formed bubble through the centring device described in \S \ref{sec:Experimental setup}. A flow rate of $Q_{i}^* = 78$~ml/min ($Q_{i} = 0.03$) was chosen so that upon exit, the bubble adopted a steadily-propagating elongated shape, which straddled the rail symmetrically about the centreline of the channel ($y_c = 0$); see \S \ref{sec:Bifurcation diagram} for a summary of the steady states of the system. Once this symmetric steady state had been reached a few centimetres downstream of the centring device, the flow was interrupted (at time $t_{i}^*$) and the bubble was left to relax its shape at $Q^*=0$. This ensured that the geometry of the centring device did not influence the evolution of the bubble. Note that the interruption of the flow led to the increase of the projected area of the bubble to its static value $A_s^*$ on a time scale much shorter than that of the bubble relaxation process.

\begin{figure}
	\centerline{\includegraphics[scale=1]{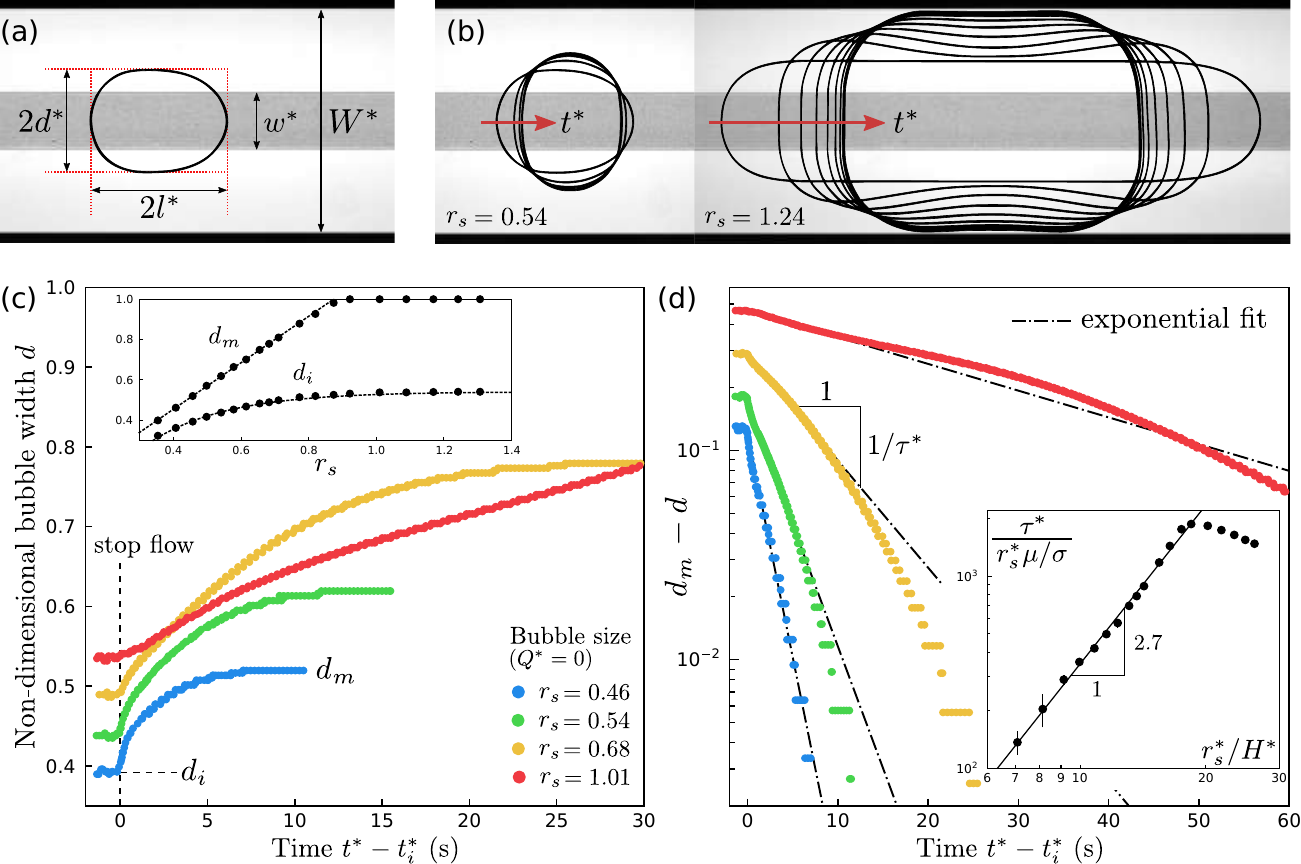}}
	\caption{Bubble relaxation at $Q^*=0$. (a) Top-view image of a bubble of width $2d^*$ and length $2l^*$, symmetric about the channel centreline, with non-dimensional static radius $r_s = 0.54$. The grey band indicates the rail. (b) Superposition of bubble shapes showing the relaxation of a bubble with $r_s = 0.54$ (left) and $r_s = 1.24$ (right). The initial slender symmetric shape corresponds to the stable symmetric propagation mode at $Q_{i}^* = 78$~ml/min which was reached before the flow was interrupted at $t^* = t_{i}^*$. The time steps between two successive images are $\Delta t^* = 3.2$~s (left) and $\Delta t^* = 26.7$~s (right). (c) Time evolution of the non-dimensional bubble width $d = 2d^*/W^*$ during relaxation from $d_{i}$ at $t^* = t_{i}^*$ to a maximum value $d_{m}$ for different bubble sizes. Inset: $d_{i}$ and $d_{m}$ plotted as a function of $r_s$. Empirical fits: $d_{m} = 1.15 r_s$ if $r_s < 0.87$ and $d_{m} = 1$ if $r_s > 0.87$; $d_{i} = 0.54 - 1.05 \exp(-4.35 r_s)$. (d) Variation of $d_{m} - d$ as a function of $t^* - t^*_i$ plotted on a semi-log scale. The initial regime is fitted by an exponential from which we measure a characteristic relaxation time $\tau^*$. Inset: $\tau^* / (r_s^* \mu / \sigma)$ against confinement ratio $r_s^*/H^*$ fitted with a power law of the form $y = 0.63 x^{2.7}$.}
	\label{fig:reshape}
\end{figure}

The bubble relaxation process is shown in figure \ref{fig:reshape}. We characterise the shape of the bubble by its maximum length $2l^*$ and maximum width $2d^*$ aligned along $x^*$ and $y^*$, respectively. These were measured by identifying the smallest rectangle bounding the bubble contour; see figure \ref{fig:reshape}(a). They are related through the range of shapes adopted by the bubble during relaxation and thus, we choose to parametrize the bubble shape solely by its non-dimensional width $d = 2d^*/W^*$. 

Typical shape evolutions are shown in figure \ref{fig:reshape}(b), where images taken at a constant time interval are superposed for bubbles of static size $r_s = (2 \sqrt{A_s^*/ \upi}) / W^*=0.54$ (left) and $r_s = 1.24$ (right). Starting from an elongated initial bubble shape at $t_{i}^*$, the length of the bubble decreases and its non-dimensional width $d$ increases from an initial value $d_{i}$ at time $t_i^*$ to a plateau value $d_{m}$, as shown in figure \ref{fig:reshape}(c) for the range of bubbles sizes investigated. During relaxation, the bubble area decreased by $\sim 2$~\% because the fraction of the bubble situated over the rail decreased. Values of $d_{i}$ and $d_{m}$ are plotted against $r_s$ in the inset of figure \ref{fig:reshape}(c), which indicates that the bubble relaxes to the width of the channel for $r_s>0.87$. In the bubble propagation experiments in \S \ref{sec:observations}, we are limited to initial bubble widths ranging between $d_{i}$ and $d_{m}$ for a given bubble size and to the shapes explored by the bubble during relaxation. We used the curves shown in figure \ref{fig:reshape}(c) to infer the time required to reach a desired value of $d$ from which to initiate the propagation of the bubble.

The time evolution of $d_{m} - d$ is plotted on a semi-log scale in figure \ref{fig:reshape}(d), revealing that the relaxation of small bubbles ($r_s \le 0.46$) is approximately exponential. Larger bubbles (e.g. $r_s =1.24$) exhibit qualitatively different relaxation, which is only exponential at early times, presumably because of the influence of the side walls of the channel. The exponential part of these relaxation curves was fitted to obtain a characteristic relaxation time scale $\tau^*$, which saturates to approximately $35$~s for bubbles of size $r_s > 1$. The non-dimensional decay time $\tau^* \sigma/ (r_s^* \mu)$ is plotted against the bubble confinement ratio $r_s^*/H^*$ in the inset of figure \ref{fig:reshape}(d). It follows a power law of exponent $2.7$ for bubbles smaller than the width of the channel ($r_s < 1$), which is larger than the exponent $2.0$ found by \citet{brun2013generic} for the exponential relaxation of elliptical bubbles towards a circular shape in a liquid-filled, infinite Hele-Shaw cell of uniform depth, in the absence of external flow. A possible explanation is that the relaxation process is accelerated by the presence of the rail which supports pressure gradients driving the bubble into the deeper parts of the channel.  

Due to the presence of the rail, small bubbles systematically migrated towards one of the deeper parts of the channel and reached an asymmetric equilibrium state, consistent with the results of \citet{franco2018bubble} for $r\le 0.87$, whereas sufficiently large bubbles ($d_{m} \sim 1$) tended to a symmetric equilibrium state. This migration was associated with a decrease in static bubble area of up to 7~\% as the fraction of bubble volume residing in the deeper parts of the channel increased. Small bubbles always migrated to the same side of the rail because of a small unavoidable bias in the levelling of channel. When this bias was minimised, the migration time scale was significantly longer than the relaxation time scale (typically one minute compared to $10$~s for $r_s = 0.54$), which ensured that the bubble had a centroid offset of $|y_c| \leq 0.005$ when the flow was initiated. 


\section{Transient evolution of a centred bubble propagating from rest}
\label{sec:observations}

Having established how to reproducibly prepare initially centred bubbles of different shapes, we propagate each bubble along the channel by imposing a constant flow rate. We investigate the evolution of these bubbles as a function of initial bubble width and flow rate. Unless otherwise specified, we focus on bubbles sizes $r_Q = 0.54$ measured during propagation. 

\subsection{Influence of the initial bubble shape}
\label{sec:Influence of initial bubble shape}

\begin{figure}
	\centerline{\includegraphics[scale=1]{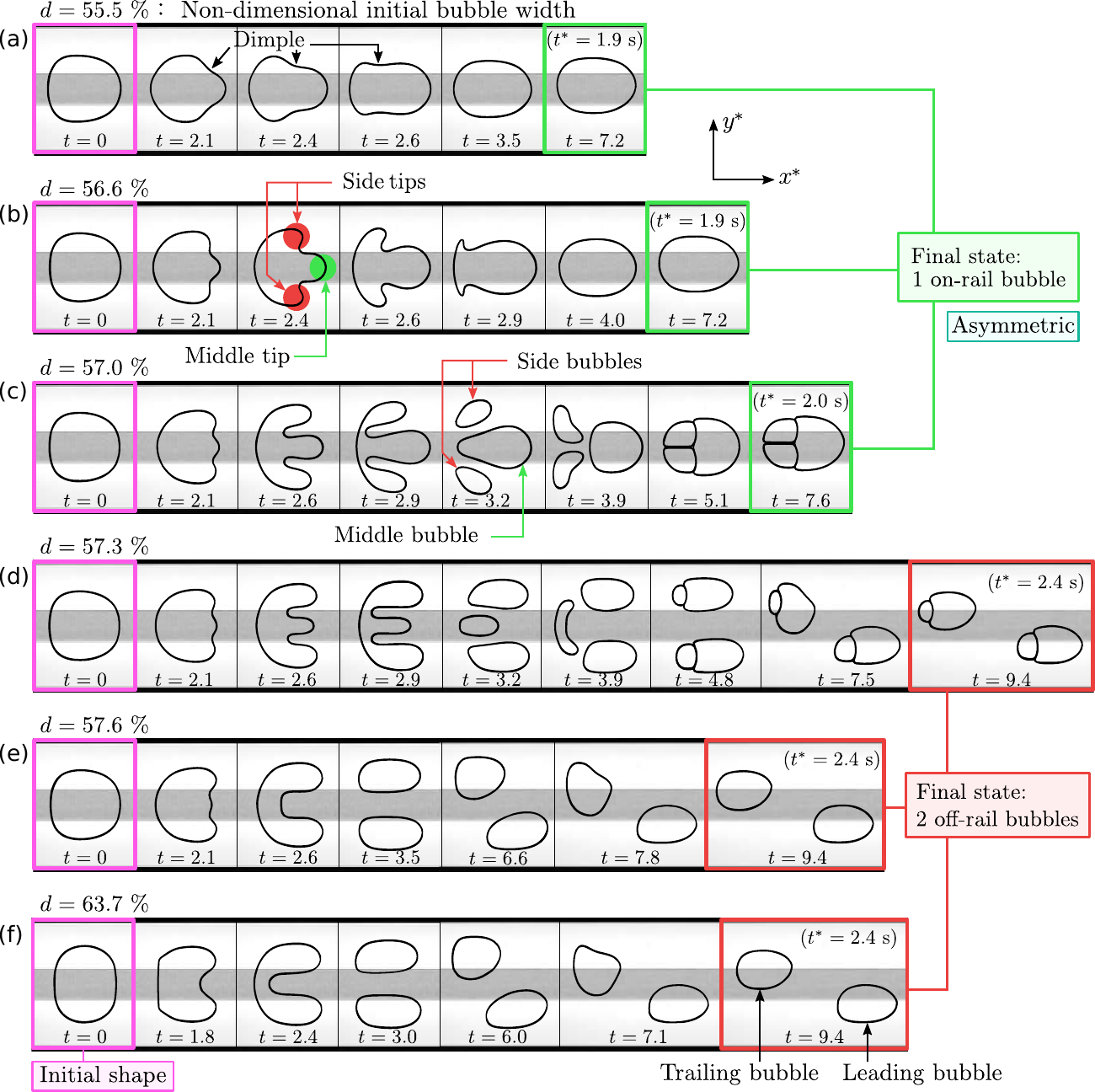}}
	\caption{Time evolution of bubbles of different initial shapes propagating from rest at a flow rate $Q^* = 186$~mL/min ($Q = 0.07$) imposed at $t^*=0$. Each row of top-view images corresponds to a time sequence with a different value of non-dimensional initial bubble width $d = 2d^*/W^*$ in the range $55.5\% \le d \le 63.7\%$. The grey band in each image indicates the rail. The non-dimensional time $t = 2 U_0^* t^*/ W^*$ elapsed since flow initiation is indicated in each snapshot, where $U_0^* = Q^*/(W^* H^*)$. The dimensional time $t^*$ is indicated in the last snapshot of each time sequence. The in-flow bubble size is $r_Q = 0.54$ (after flow-induced compression) and $d$ is measured at $t = 0$ (just before flow initiation).}
	\label{fig:map_horizontal}
\end{figure}

Figure \ref{fig:map_horizontal} shows the evolution of bubbles propagating from rest at a flow rate $Q^* = 186$~mL/min ($Q = 0.07$), with non-dimensional time elapsed since flow initiation, $t = 2 U_0^* t^*/ W^*$. Each row of images corresponds to a time sequence of bubble evolution of an initial bubble shape quantified by its non-dimensional width $d$ measured at $t = 0$ (just before flow initiation) which ranges from $55.5$~\% (a) to $63.7$~\% (f). 

For all initial bubble shapes, the long-term outcome falls into only two categories: (i) a single, steadily-propagating bubble straddling the rail (final frames in figure \ref{fig:map_horizontal}(a--c)); or (ii) two bubbles on opposite sides of the rail each moving at a constant velocity, where the larger, leading bubble propagates faster than the smaller, trailing one so that their relative distance increases in time (final frames in figure \ref{fig:map_horizontal}(d--e)). For $Q=0.07$, the single bubble straddling the rail at late times in figure \ref{fig:map_horizontal}(a--c) is asymmetric about the centreline of the channel, and this state will henceforth be referred to as `on-rail, asymmetric'. The two bubbles propagating on opposite sides of the rail at late times in figure \ref{fig:map_horizontal}(d--e) are termed `off-rail' bubbles because they do no straddle the rail, only partially overlapping it from one side.

These long-term outcomes are associated with an organised set of transient evolutions that are determined by the initial deformation of the centred bubble into single, triple or double-tipped bubbles as $d$ increases (see second column of snapshots in figure \ref{fig:map_horizontal}). Following this initial deformation, the bubble evolution may feature complex bubble break up and aggregation events. The wide range of transient evolutions obtained for relatively modest changes in $d$ ($55.5 \% \le d \le 63.7 \%$) is fully reproducible. In some cases, the final state features compound bubbles resulting from the aggregation of two or more bubbles (figure \ref{fig:map_horizontal}(c, d)) which persist until reaching the end of the channel at $Q = 0.07$. However, slow drainage of the thin oil film separating the different bubbles will eventually result in their coalescence into simple bubbles.

The simplest evolution scenarios towards single and two-bubble final states occur for the most slender (figure \ref{fig:map_horizontal}(a)) and widest (figure \ref{fig:map_horizontal}(f)) initial bubble shapes, respectively. In figure \ref{fig:map_horizontal}(a), the centred bubble deforms symmetrically about the channel centreline until $t = 3.5$, after which it migrates sideways to reach the asymmetric, on-rail state at late times ($t=7.2$). 
When the flow is initiated, the front of the bubble located over the rail is pulled forward. By $t=2.1$, dimples on the interface (i.e. regions of negative curvature) have formed at the edges of the rail. These dimples are subsequently advected towards the rear part of the bubble (in the frame of reference of the moving bubble) where they vanish shortly after $t=2.6$, so that an approximately symmetric, elongated bubble with entirely positive curvature emerges at $t=3.5$.  

In figure \ref{fig:map_horizontal}(f), the bubble initially features a wide front that encroaches into the deeper regions of the channel on either side of the rail. When flow is initiated, 
two fingers form in the deeper side-channels separated by a dimple of negative curvature over the rail ($t=1.8$), which is essentially the reverse of the initial deformation in figure \ref{fig:map_horizontal}(a). The dimple deepens into a cleft ($t=2.4$) which continues to grow as the fingers lengthen, until the bubble breaks in two ($t=3.0$). Because of the small, unavoidable asymmetry in the initial bubble shape discussed in \S \ref{sec:Initial bubble shape} (here, centroid offset $y_c = -0.005$ at $t=0$), the two fingers do not lengthen at precisely the same rate and thus, the newly formed bubbles propagating on opposite sides of the rail have slightly different sizes (the projected area of the bottom bubble is $5$ \% larger than that of the top one). We find that larger bubbles always propagate faster than smaller ones within the range of flow rates and bubbles sizes investigated experimentally and we quantify this effect in \S \ref{sec:Bifurcation diagram}. In figure \ref{fig:map_horizontal}(f), the larger bubble has visibly moved ahead of the smaller one by $t=6.0$ and the two bubbles continue to separate, which indicates that the change in topology is permanent. Both bubbles propagate steadily by $t=9.4$, with the leading bubble propagating $2$~\% faster than the trailing one.

Prior to flow initiation, the bubble is characterised by a capillary-static pressure distribution which depends on the local in-plane curvature of the bubble and depth of the channel through the Young--Laplace relation. If the bubble is left to relax, local pressure gradients generate flows in the viscous fluid which act to broaden slender bubbles as shown in \S \ref{sec:Initial bubble shape}. However, the imposition of a flow rate driving the bubble forward tends to override this initial capillary-static distribution. Each point on the bubble interface is displaced along its normal direction and thus, its normal velocity is largest when the interface is orthogonal to the direction of imposed flow along $x$. In addition, the presence of the rail means that the mobility of the fluid is reduced over the rail compared with the deeper side channels. Hence, if a portion of the bubble front encroaches into the deeper side channels, these regions of interface can advance more rapidly than the part overlapping the rail, provided that these portions are oriented so that their normal spans a small enough angle from the $x$-direction. In order for bulges to form, these portions of interface also need to be sufficiently wide to overcome capillary forces, as is the case in figure \ref{fig:map_horizontal}(f). If these criteria are not fulfilled, the portion of interface over the rail will advance most rapidly to form a central  bulge as shown in figure \ref{fig:map_horizontal}(a).

For initial bubble shapes of widths intermediate between the slender and wide bubbles shown in figures \ref{fig:map_horizontal}(a) and (f), the front of the bubble initially deforms into a shape that is hybrid between those described above, i.e., by simultaneously developing a central bulge as well as bulges in the deeper side channels. 
This intermediate initial bubble deformation is associated with the most complex bubble evolutions, which occur within a narrow range of initial shapes, $(56.8 \pm 0.1)\% \le b \le (57.5 \pm 0.1)\%$. We refer to the developing bulges as `side' and `middle' tips as labelled in figure \ref{fig:map_horizontal}(b) ($t=2.4$). As $d$ is increased, the growth rate of the side tips increases relative to that of the middle tip, which is reduced ($t=2.1-2.6$ in figure \ref{fig:map_horizontal}(b--e)). The competition between side and middle tips is a key factor in the evolution of the bubble. In figure \ref{fig:map_horizontal}(b), the middle tip grows and the side tips retract, so that by $t=4.0$, the bubble evolves similarly to that in figure \ref{fig:map_horizontal}(a). By figure \ref{fig:map_horizontal}(c), the rate of growth of the side tips has increased and is only marginally smaller than that the middle tip, resulting in a break up into three bubbles ($t=3.2$). The larger middle bubble, which is approximately centred on the rail, pulls the smaller side bubbles in behind it where pressure is lower ($t=3.9$) to form a centred compound bubble ($t=5.1$), which finally drifts sideways to reach an asymmetric, on-rail state ($t=7.6$). 

The transition in long-term behaviour, from a single on-rail bubble to two off-rail bubbles, occurs between figures \ref{fig:map_horizontal}(c) and \ref{fig:map_horizontal}(d) and coincides with the side tips overtaking the middle tip prior to bubble break up. In figure \ref{fig:map_horizontal}(d), this leads to a break up into two side bubbles propagating ahead of a smaller centred bubble ($t=3.2$). This smaller bubble is stretched across the rail as it is attracted similarly towards both side bubbles ($t=3.9$), so that in turn, it breaks into two small bubbles which are pulled in behind the side bubbles ($t=4.8$). In figure \ref{fig:map_horizontal}(e) the side tips have become dominant so that the bubble breaks into two parts. In both cases, the remaining two-bubble evolution is similar to that in figure \ref{fig:map_horizontal}(f). 


\subsection{Influence of the flow rate}
\label{sec:Influence of the flow rate}

We now turn to the influence of the flow rate on bubble evolution from a prescribed initial condition and successively examine the limits of slender and wide initial bubbles. 

\subsubsection{Initially slender bubbles}
\label{sec:For initially slender bubbles}

\begin{figure}
	\centerline{\includegraphics[scale=1]{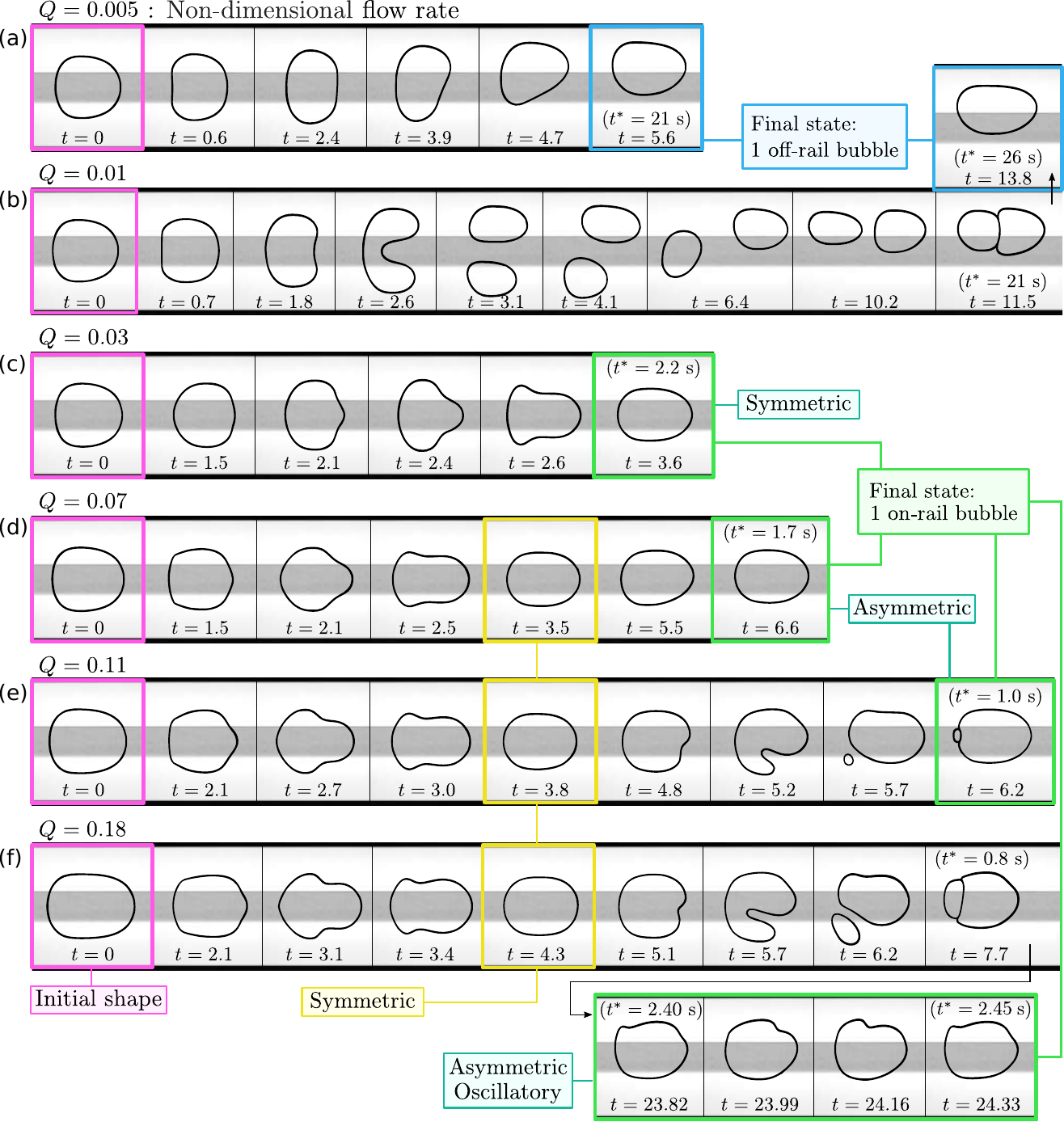}}
	\caption{Time evolution of initially slender bubbles of non-dimensional width $53\; \% \le d \le 55\; \%$ with $r_Q=0.54$, where each row of top-view images corresponds to a time sequence at a different value of the flow rate imposed at $t^* = 0$ from $Q^* = 13$~mL/min ($Q = 0.005$) to $Q^* = 477$~mL/min ($Q = 0.18$). The time labelling is the same as in figure 4.}
	\label{fig:map_vertical}
\end{figure}

Figure \ref{fig:map_vertical} shows the evolution of initially slender bubbles (within a narrow range of widths $53\;\% \le d \le 55\; \%$) as a function of flow rate, where each row of images corresponds to a time sequence from the instant $t = 0$ when a fixed value of the flow rate was imposed in the range $0.005 \le Q \le 0.18$. In order to maintain the in-flow bubble size at a fixed value $r_Q = 0.54$, the size of the bubble before flow-induced compression needs to increase with $Q$; see the initial shapes at $t=0$ in the column of magenta frames in figure \ref{fig:map_vertical} where the bubble length increases while $d$ remains approximately constant. 

For all values of $Q$, the long-term outcome is a single (simple or compound) bubble which is stable and propagates steadily. We observe three types of steadily-propagating, invariant bubbles: strongly asymmetric bubbles, which partially overlap the rail from one side, referred to as `off-rail' (final frames in figure \ref{fig:map_vertical}(a, b)), a bubble which straddles the rail symmetrically about the channel centreline, referred to as `on-rail, symmetric' (final frame in figure \ref{fig:map_vertical}(c)) and `on-rail, asymmetric' bubbles (final frames in figure \ref{fig:map_vertical}(d, e)). In figure \ref{fig:map_vertical}(f), the bubble exhibits periodic oscillations with the green frames showing one period. This state is referred to as `on-rail, asymmetric, oscillatory' and is related to previously studied bubble and finger oscillations induced by channel depth perturbations \citep{pailha2012oscillatory,jisiou2014geometry,thompson2014multiple}.
It features the advection of interface dimples along the side of the bubble that is furthest from the channel centreline (top region in the final frames of figure \ref{fig:map_vertical}(f)). These dimples are periodically generated at the point where the front of the bubble meets the edge of the rail. 

All bubbles in figure \ref{fig:map_vertical} initially deform symmetrically about the centreline of the channel. This key feature of the bubble evolution will be interpreted in terms of the edge states of the one-bubble system in \S \ref{sec:single-bubble unstable solutions}. For the smallest values of $Q$ (figure \ref{fig:map_vertical}(a,~b)), the bubble initially broadens in response to the imposed flow because surface tension forces are dominant and the bubble evolves as though relaxing to static equilibrium. This results in the bubble migrating off the rail without breaking for $Q=0.005$, whereas for $Q=0.01$, the decreasing influence of surface tension is associated with the growth of a dimple of negative curvature on the front part of the bubble ($t=1.8$), which leads to break up into two bubbles on opposite sides of the rail ($t=3.1$). These have different sizes because of the small, unavoidable asymmetry in the original bubble shape (centroid offset $y_c = 0.002$ at $t=0$). The smaller trailing bubble migrates across the rail ($t=6.4$) and is pulled in behind the larger leading bubble resulting in an off-rail compound bubble ($t=11.5$). In contrast with figure \ref{fig:map_horizontal}(c, d), we observe coalescence into a simple bubble before the bubble reaches the end of the channel ($t=13.8$) because the oil film separating the bubbles has sufficient time to drain due to the smaller flow rate. The bubble velocity is the same before and after coalescence. 

As $Q$ increases, the early-time broadening of the bubble promoted at low $Q$ is progressively replaced by the growth of a bulge over the rail. This is illustrated in figure \ref{fig:map_vertical}(c--f) which shows transient bubble evolutions similar to figure \ref{fig:map_horizontal}(a) until the bubble reaches an elongated, symmetric shape, see, e.g., final frame in figure \ref{fig:map_vertical}(c) and yellow frames in figure \ref{fig:map_vertical}(d--f). 
We find that the next stage of evolution may involve break up and aggregation events following tip-splitting beyond a threshold value of the flow rate $Q_{ts} = 0.085 \pm 0.005$. This is illustrated in figure \ref{fig:map_vertical}(e, f), where the bubble breaks into two parts of widely different sizes which subsequently aggregate. These events are increasingly likely the larger the flow rate and we refer to \S \ref{sec:High flow rates} for a discussion of this late stage of evolution.

Note that the compression of the bubble occurs gradually over increasing time intervals as flow rate increases, so that in figure \ref{fig:map_vertical}(f) ($Q=0.18$) for example, an approximately constant projected area is only reached for $t>4.0$. We provide evidence in Appendix \ref{sec:Influence of the bubble size} that increasing the size of the bubble beyond $r_Q=0.54$ does not influence its evolution significantly and thus, the bubble evolutions presented in figure \ref{fig:map_vertical} are not expected to differ from the idealised case of a bubble of fixed volume. 


\subsubsection{Initially wide bubbles}
\label{sec:For initially wide bubbles}

\begin{figure}
	\centerline{\includegraphics[scale=1]{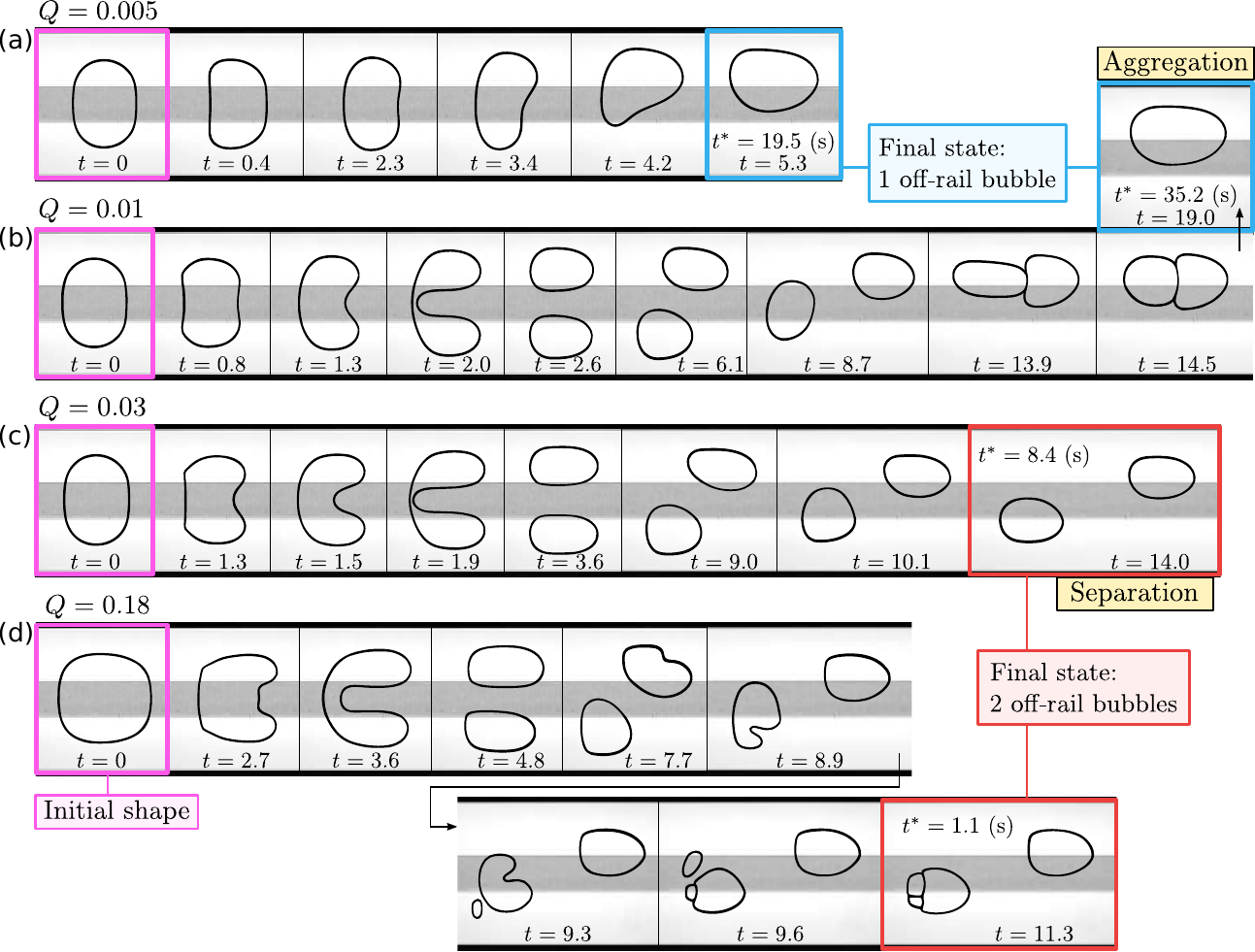}}
	\caption{Time evolution of initially wide bubbles with non-dimensional width $d = 63.4$~\% and $r_Q=0.54$. Each row of top-view images corresponds to a time sequence at a different value of the flow rate imposed at $t^* = 0$ between $Q^* = 13$~mL/min ($Q = 0.005$) and $Q^* = 477$~mL/min ($Q = 0.18$). The time labelling is the same as in figure 4.}
	\label{fig:map_vertical_wide}
\end{figure}

Figure \ref{fig:map_vertical_wide} shows the evolution of initially wide bubbles ($d = 63.4$~\%) as a function of the flow rate in the range $0.005 \le Q \le 0.18$ with a layout similar to figure \ref{fig:map_vertical}.  
For low flow rates (figure \ref{fig:map_vertical_wide}(a, b)), the bubble evolution is similar to that of the slender bubbles in figure \ref{fig:map_vertical}(a, b), resulting in a steadily-propagating off-rail bubble. This suggests that the bubble evolution is insensitive to the initial shape of the bubble within this range of flow rates. 

However, for higher flow rates (figure \ref{fig:map_vertical_wide}(c, d)), the long-term outcome is two off-rail bubbles, where the leading bubble is larger than the trailing one so that their relative distance increases with time. The early-time evolution is similar to that of figure \ref{fig:map_vertical_wide}(b), in that the bubble breaks into two parts of similar but not identical sizes. Whereas in figure \ref{fig:map_vertical_wide}(b), the smaller trailing bubble migrates across the rail and aggregates with the larger leading bubble, in figure \ref{fig:map_vertical_wide}(c), the two bubbles remain on opposite sides of the rail and separate, like in figure \ref{fig:map_horizontal}(f). The transition between the time evolutions leading to aggregation (figure \ref{fig:map_vertical_wide}(b)) and separation (figure \ref{fig:map_vertical_wide}(c)) occurs at a threshold value of the flow rate $Q_c = 0.017 \pm 0.001$. We will show in \S \ref{sec:Interaction between two bubbles of comparable sizes} how these distinct evolutions are linked to the presence of two-bubble edge states. 


In figure \ref{fig:map_vertical_wide}(c), the smaller bubble exhibits transient oscillations (at around $t=10.1$) before reaching a state of steady propagation. As the flow rate increases, these oscillations increase in amplitude and promote the repeated break up of the smaller trailing bubble, as shown in figure \ref{fig:map_vertical_wide}(d) where tiny bubbles created through break up aggregate with the bubble that spawned them. These secondary break up events occur for $Q>0.135 \pm 0.005$ and increase in complexity and unpredictability as $Q$ increases, leading to complex final states involving more than two bubbles for the highest flow rates investigated.

\subsection{Summary of bubble evolution}
\label{sec:Summary of bubble evolution}

We now proceed to summarise the transient bubble evolutions reported in \S \ref{sec:Influence of initial bubble shape} and \S \ref{sec:Influence of the flow rate} in a phase diagram and map the different long-term behaviours of single bubbles onto a bifurcation diagram.

\subsubsection{Phase diagram of bubble evolution}
\label{sec:Phase diagram of bubble evolution}

\begin{figure}
	\centerline{\includegraphics[scale=1]{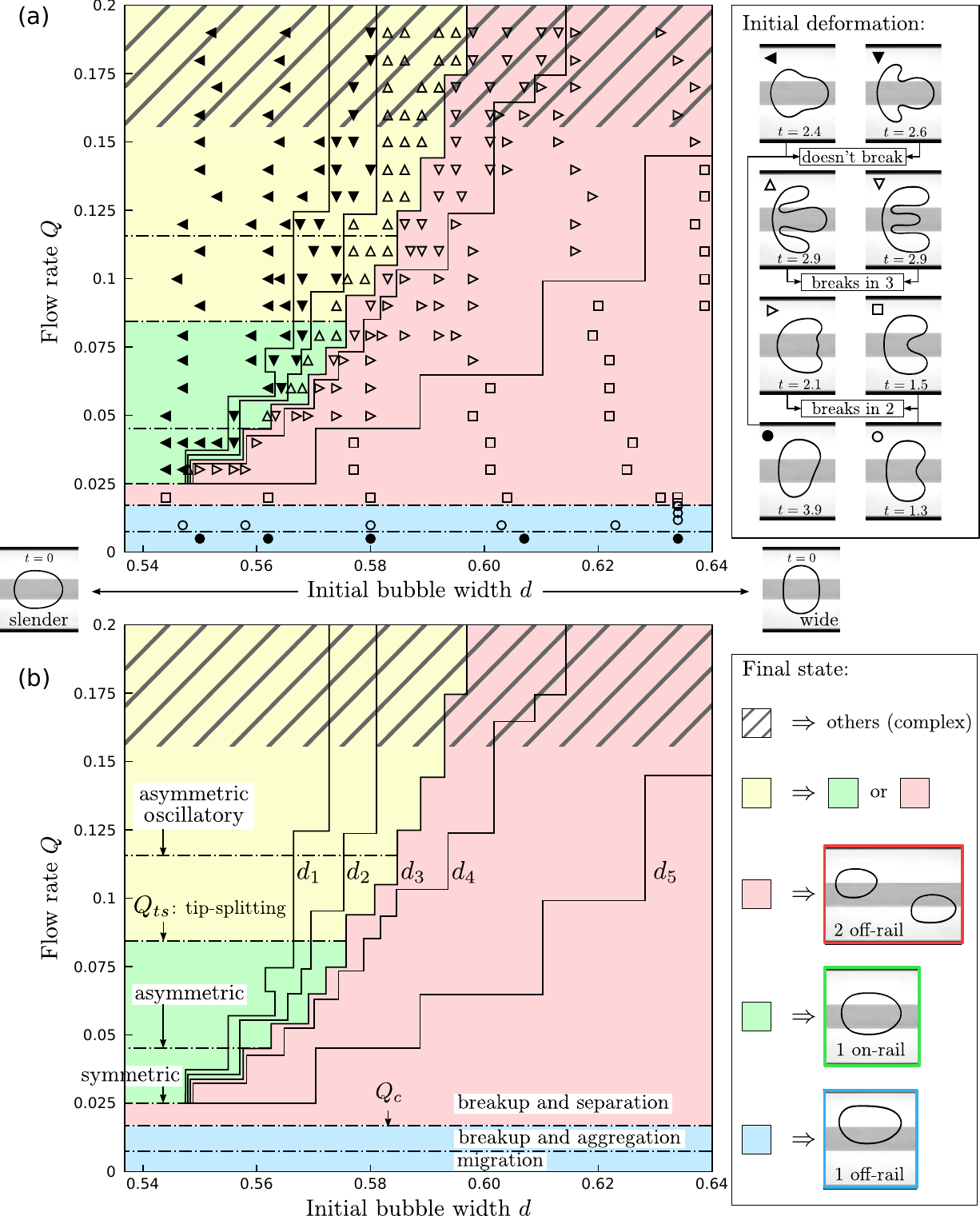}}
	\caption{(a) Phase diagram of the different time evolutions of a bubble with $r_Q = 0.54$ in terms of its initial non-dimensional width $d$ and the non-dimensional flow rate $Q$. (b) Reproduction of (a) without the data points to allow for annotations. We distinguish eight different types of transient evolution, each denoted by a symbol illustrated by a snapshot of the associated initial deformation on the right-hand side of the diagram in (a). The piecewise linear solid lines labelled $d_i$ ($i=1-5$) in (b) indicate approximate boundaries between different regions. The background colour indicates the long-term outcome of the bubble evolution as illustrated in the legend on the right-hand side of the diagram in (b). The hatching at high flow rate highlights the increasing diversity of final outcomes as $Q$ increases (see figure \ref{fig:map_tristability}). Figure \ref{fig:map_horizontal} is representative of the bubble evolution along a horizontal line of constant $Q$ for $Q > 0.025 \pm 0.005$, while figures \ref{fig:map_vertical} and \ref{fig:map_vertical_wide} correspond to the behaviour along vertical lines of constant $d$ for slender ($d<d_3$) and wide ($d>d_3$) initial bubbles, respectively.}
	\label{fig:mapexp}
\end{figure}


The influence of the initial bubble width and the flow rate is summarised in a phase diagram in figure \ref{fig:mapexp}(a), where we classify more than two hundred experimental time evolutions into eight categories, each denoted by a different symbol. Each type of transient evolution occurs in a simply connected region of the diagram. The diagram is reproduced in figure \ref{fig:mapexp}(b) without the data points to allow space for annotations. 

For $Q \le 0.025 \pm 0.005$, the evolution of the bubble does not depend on initial conditions, and three different types of evolution are observed. When $Q > 0.025 \pm 0.005$, the variation of the initial bubble width yields six distinct types of early-time evolution, previously introduced in figure \ref{fig:map_horizontal}. 
The rich variety of transient evolutions is concentrated over a small range of initial bubble widths for moderate flow rates above this threshold, and the ranges of $d$ for which these initial evolutions are observed widen as $Q$ increases.

The long-term outcomes are indicated by the background colour: blue for single off-rail bubbles, green for single on-rail bubbles (symmetric, asymmetric or oscillating) and red for two off-rail bubbles (or occasionally three as will be shown in \S \ref{sec:Moderate flow rates}). The yellow region corresponds to bubbles which may undergo tip-splitting following their early-time symmetric evolution for $Q>Q_{ts}$, resulting in either a single on-rail bubble or two off-rail bubbles; see \S \ref{sec:High flow rates}. The hatching highlights the increasing diversity of final outcomes (with possibly more than two bubbles) as the flow rate increases; see \S \ref{sec:High flow rates}.

For all initial bubble widths investigated, the long-term behaviour of the bubble transitions from one off-rail bubble to two off-rail bubbles at a flow rate $Q_c =  0.017 \pm 0.001$. For $Q < Q_c$, the single off-rail bubble is reached by migrating off the rail without breaking ({\large $\bullet$}) or by breaking into two parts which subsequently aggregate ({\large $\circ$}). By contrast, the two bubbles separate after break up for $Q_c < Q \le 0.25 \pm 0.005$ ({\tiny $\square$}).

For $Q > 0.025 \pm 0.005$, the boundary $d_3$ separates the red region, where the long-term outcome is always two (or three) off-rail separating bubbles, from the green and yellow regions where a single on-rail bubble is a possible long-term outcome. In the green and yellow regions, dash-dotted lines in figure \ref{fig:mapexp}(b) indicate the critical values of flow rate at which different single on-rail bubble invariant states emerge. These are detailed in an experimental bifurcation diagram in \S \ref{sec:Bifurcation diagram}. 

The boundary $d_3$ corresponds to the initial bubble width for which side and middle tips of the three-tipped transient bubbles introduced in figure \ref{fig:map_horizontal} grow at the same rate during early-time evolution. For $d_1 \le d \le d_3$ the middle tip grows faster than the side tips whereas the converse is true for $d_3 \le d \le d_5$. For $d<d_2$, the bubble does not break following initial deformation, whereas for $d_2<d<d_4$ it breaks in three ({\small $\bigtriangleup$} and {\small $\bigtriangledown$}) and for $d>d_4$ it breaks in two. The value $d_1$ indicates the development of side tips (the maximum number of intersections of the bubble contour with a line perpendicular to the side walls increases from two ($\blacktriangleleft$) to six ($\blacktriangledown$)). Similarly, the value $d_5$ indicates the disappearance of the middle tip (the maximum number of intersections of the bubble contour with a line perpendicular to the side walls decreases from six ($\vartriangleright$) to four ({\tiny $\square$})).

\subsubsection{Single-bubble invariant modes of propagation}
\label{sec:Bifurcation diagram}

\begin{figure}
	\centerline{\includegraphics[scale=1]{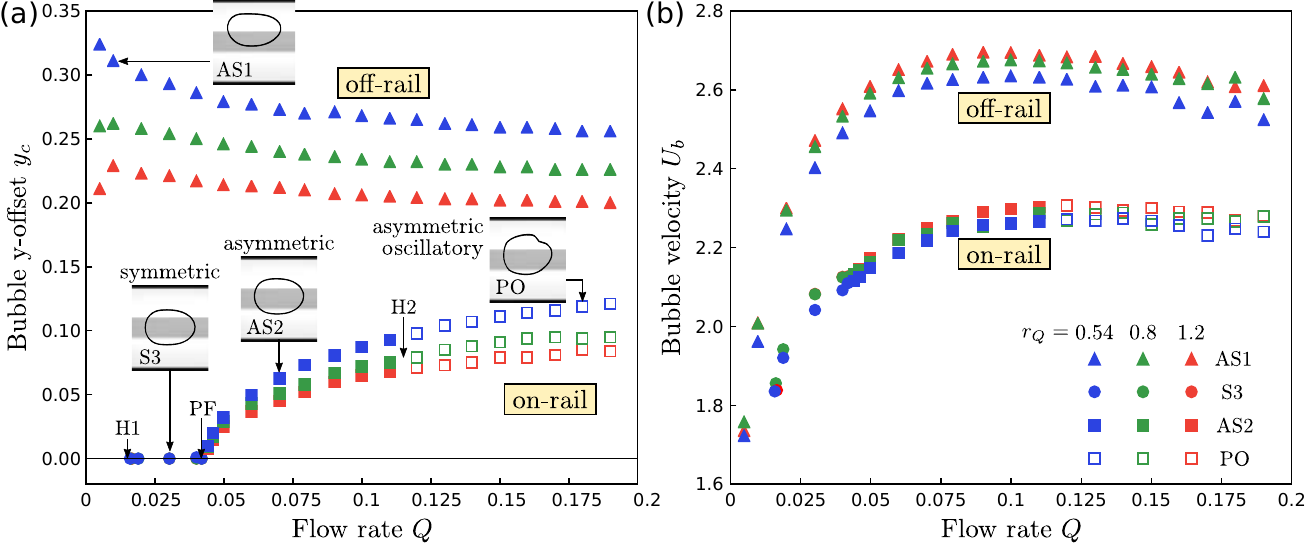}}
	\caption{Experimental bifurcation diagram where the bubble $y$-offset $y_c = y_c^*/(W^*/2)$ (a) and the bubble non-dimensional velocity $U_b = U_b^*/U_0^*$ (b) are plotted as functions of the non-dimensional flow rate $Q$ for three bubble sizes $r_Q = 0.54$, $0.8$ and $1.2$. The off-rail branch AS1 and the three on-rail branches, which are symmetric (S3) asymmetric (AS1) and oscillating asymmetrically (PO) are illustrated by the final-state snapshots of figure \ref{fig:map_vertical}(b--f), respectively. H1 indicates the first appearance of on-rail bubbles, PF and H2 refer to the bifurcations between the on-rail branches. For the periodic orbit PO, the reported values of $y_c$ and $U_b$ are averaged over one period.}
	\label{fig:bifurcation_exp}
\end{figure}

Figure \ref{fig:bifurcation_exp} shows the single-bubble invariant modes of propagation observed experimentally in terms of their non-dimensional centroid offset $y_c$ (a) and velocity $U_b$ (b), both plotted as functions of $Q$ for three values of the bubble size $r_Q$. These bifurcation diagrams extend those reported by \citet{franco2017bubble,franco2018bubble} for $Q<0.014$ to larger values of the flow rate up to $Q=0.19$. As shown in figure \ref{fig:bifurcation_exp}(a), the branch of off-rail asymmetric bubbles (AS1) extends from the capillary-static limit to the largest flow rate investigated. The AS1 mode of propagation was reached by initially positioning the bubble off-rail prior to imposing the flow rate $Q$ (following the same procedure as in \S \ref{sec:Effect of the flow on the projected area of the bubble}).

The other invariant modes of propagation are those classified as `on-rail' in figure \ref{fig:mapexp}. As shown in figure \ref{fig:bifurcation_exp}(a), the symmetric branch S3 emerges beyond a threshold flow rate $Q_{H1}$. We obtained the smallest value $Q_{H1}=0.016$ by imposing the flow while the bubble was at rest inside the centring device depicted in figure \ref{fig:setup}(c). However, the true value could be smaller because the long-term outcome is influenced by the expansion ratio of the centring device which determines the bubble shape at its exit and thereby the bubble's dynamical evolution. 
The subsequent transition to an on-rail, asymmetric state (AS2) at $Q_{PF}=0.0435$, obtained from a linear fit of centroid offset values $y_c^2(Q)$ near the transition, is consistent with an imperfect supercritical pitchfork bifurcation. Finally, the transition to an on-rail, asymmetric and oscillatory state (PO) at $Q_{H2}= 0.118$ is consistent with a supercritical Hopf bifurcation; see Appendix \ref{sec:Supercritical Hopf bifurcation} for measurements of the amplitude and frequency of these oscillations. The variation of these critical flow rates with bubble size was less than $3$~\% within the range investigated.

As shown in figure \ref{fig:bifurcation_exp}(b), off-rail bubbles propagate faster than on-rail bubbles at a given flow rate due partly to lower viscous resistance in the deeper parts of the channel. Also, bubble velocity systematically increases with bubble size over the range of bubble sizes and flow rates investigated. This is consistent with the bubble evolutions where the larger leading bubble propagates faster than the smaller trailing one; see, e.g., figure \ref{fig:map_horizontal}(f). 

\subsection{Sensitivity of the transient evolution}
\label{sec:Sensitivity of the transient evolution}

We now turn our attention to dynamical evolutions of the bubble that were particularly sensitive to either the initial condition of the system or imperfections of the rail. We distinguish situations where sensitivity occurs at moderate flow rate ($Q<Q_{ts}$; see figure \ref{fig:mapexp}) from those where sensitivity occurs at high flow rate ($Q \ge Q_{ts}$; yellow region and hatched region in figure \ref{fig:mapexp}).

\subsubsection{Moderate flow rates}
\label{sec:Moderate flow rates}

We observed that the transient evolution of the bubble exhibited enhanced sensitivity to the value of the initial offset of the bubble from the channel centreline $y_c$, for flow rates slightly larger than the value $Q_c=0.017 \pm 0.001$, which marks the transition between aggregation (figure \ref{fig:map_vertical_wide}(b)) and separation (figure \ref{fig:map_vertical_wide}(c)). We found that the final outcome of the system is influenced by the value of the size ratio of the two bubbles formed after break up, which depends on $y_c$. For example, at $Q=0.020$, the two bubbles resulting from break up only separated when the leading bubble was at most $10$~\% larger than the trailing bubble ($|y_c| < 0.002$), while they aggregated when its size exceeded that of the trailing bubble by $20$~\% to $80$~\% ($0.003 < |y_c| < 0.008$). For intermediate values, the smaller trailing bubble could stretch sufficiently across the rail to break in two, leading to a final state where three bubbles, organised in order of decreasing size and increasing velocity, separate indefinitely. For this reason, extra care was taken when estimating $Q_c$ to minimise the asymmetry of the initial bubble to offsets $|y_c|< 0.002$, which was achieved through accurate levelling of the channel in order to minimise migration of the bubble during the relaxation process described in \S \ref{sec:Initial bubble shape}.


\begin{figure}
	\centerline{\includegraphics[scale=1]{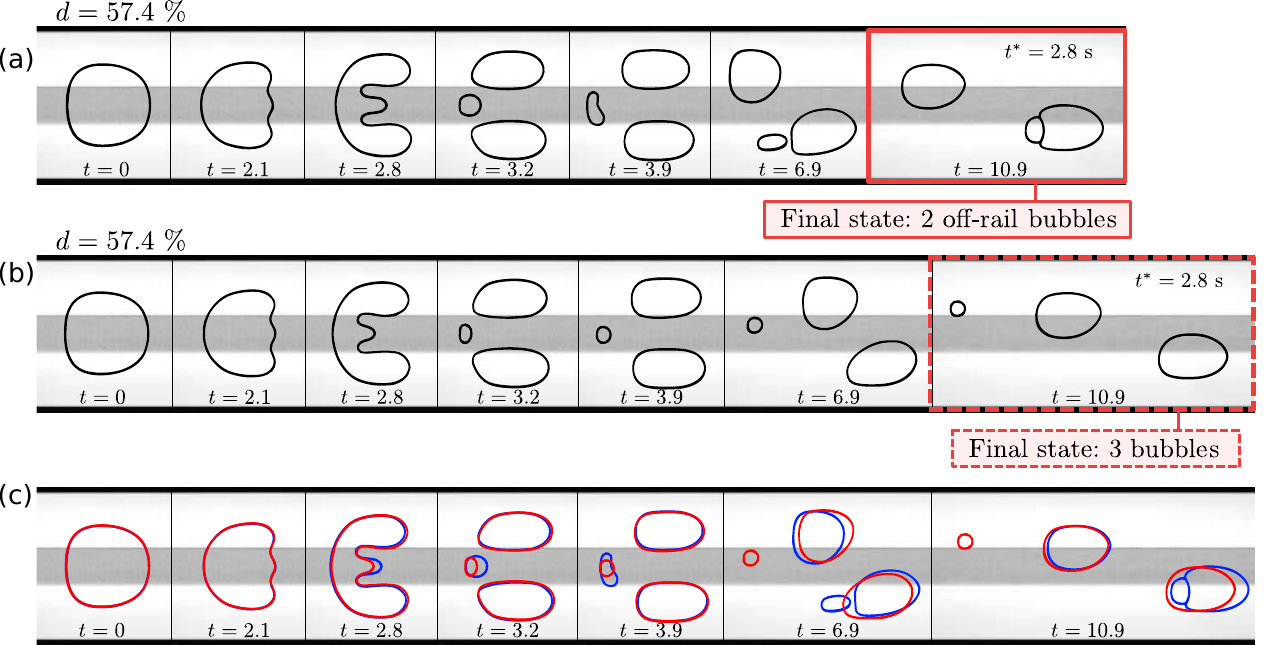}}
	\caption{(a,b) Time evolution of a bubble of initial width $d = 57.4\%$ propagating from rest at a flow rate $Q^* = 186$~mL/min ($Q = 0.07$) with $r_Q=0.54$. (c) Superposition of (a) in blue and (b) in red. Both experiments performed for the same experimental parameters feature break up into three bubbles. However, the size of the middle bubble differs visibly, which leads to different long term outcomes. The time labelling is the same as in figure 4.}
	\label{fig:map_horizontal_reproducibility}
\end{figure}

In addition, we observed that evolutions where break up events had a propensity to generate tiny bubbles featured reduced experimental reproducibility. This is illustrated in figure \ref{fig:map_horizontal_reproducibility} where successive experiments, performed at the same value of $d=57.4\%$ for $Q=0.07$, evolve towards different final outcomes involving either two or three separating bubbles. The difference between the time evolutions of figures \ref{fig:map_horizontal_reproducibility}(a) and \ref{fig:map_horizontal_reproducibility}(b) (superposed in figure \ref{fig:map_horizontal_reproducibility}(c)) is caused by differences in relative bubble sizes after break up, stemming from small unavoidable differences in the initial bubble shape of the order of $0.5$~\% in area ($t=0$), which is the typical reproducibility limit of the experimental setup.

\subsubsection{High flow rates}
\label{sec:High flow rates}

\begin{figure}
	\centerline{\includegraphics[scale=1]{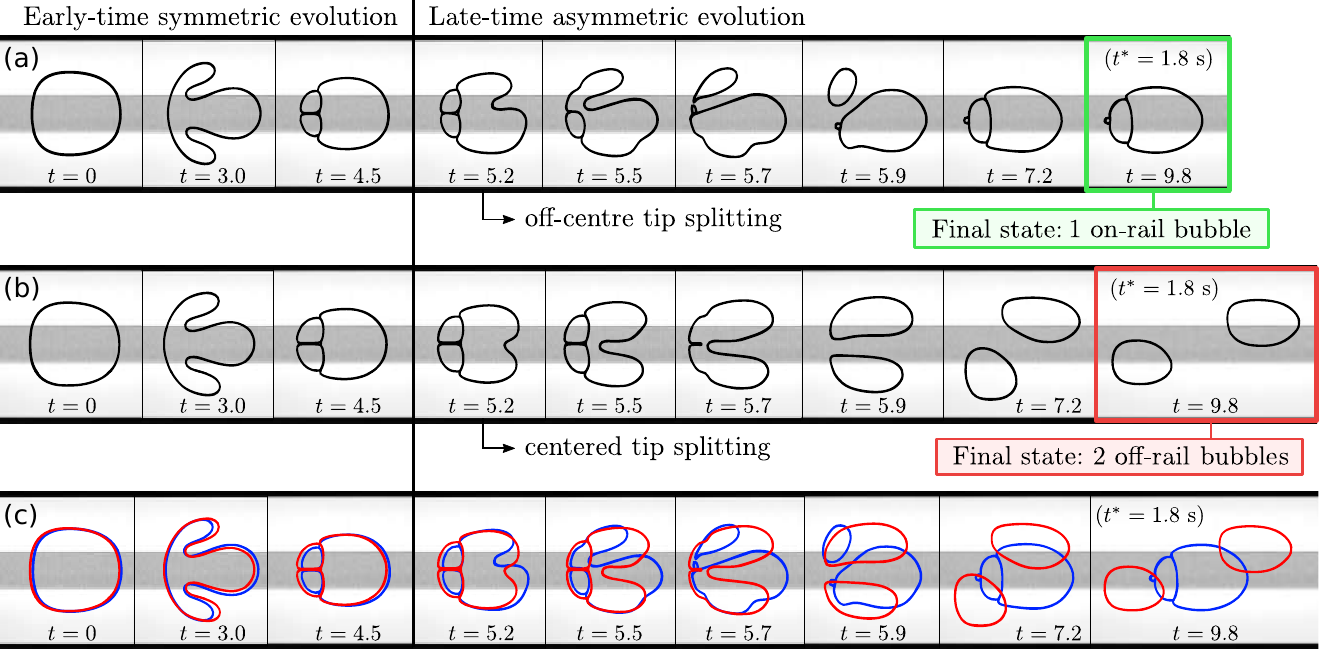}}
	\caption{(a,b) Evolution of a bubble of initial width $d = 57.7$~\% propagating from rest at a flow rate $Q^* = 265$~mL/min ($Q = 0.10$) with $r_Q=0.54$. (c) Superposition of (a) in blue and (b) in red. Both experiments performed for the same experimental parameters lead to the formation of an approximately symmetric transient compound bubble ($t=4.5$), which either undergoes off-centre (a) or centred (b) tip-splitting leading to different final outcomes. The time labelling is the same as in figure 4.}
	\label{fig:bug_break}
\end{figure}

We mentioned in \S \ref{sec:For initially slender bubbles} that initially slender bubbles may break up via tip-splitting for flow rates beyond the threshold value $Q_{ts}= 0.085 \pm 0.005$, following a reproducible symmetric early-time evolution. This second stage of bubble evolution is not reproducible in the experiment, as illustrated in figure \ref{fig:bug_break} where two successive experiments at the same parameters evolve towards radically different final outcomes. 

The early-time symmetric evolution, which here involves break up and aggregation, leads to an approximately symmetric compound bubble by $t=4.5$. In both figures \ref{fig:bug_break}(a) and (b), a dimple forms at the bubble tip ($t=5.2$), but at different locations (more centrally in (b) than in (a)). This results in a second break up which produces two bubbles of similar sizes in figure \ref{fig:bug_break}(b) and widely different sizes in figure \ref{fig:bug_break}(a) ($t=5.9$). In the latter case, the system then evolves as in figure \ref{fig:map_vertical}(e), resulting in a steadily-propagating on-rail, asymmetric bubble ($t=9.8$). However, the two bubbles in figure \ref{fig:bug_break}(b) ultimately propagate on opposite sides of the rail ($t=9.8$), with the larger leading bubble propagating faster so that they separate with increasing time. This indicates that the position of the dimple relative to the bubble tip is a key factor in predicting the long-term evolution of the system.

\begin{figure}
	\centerline{\includegraphics[scale=1]{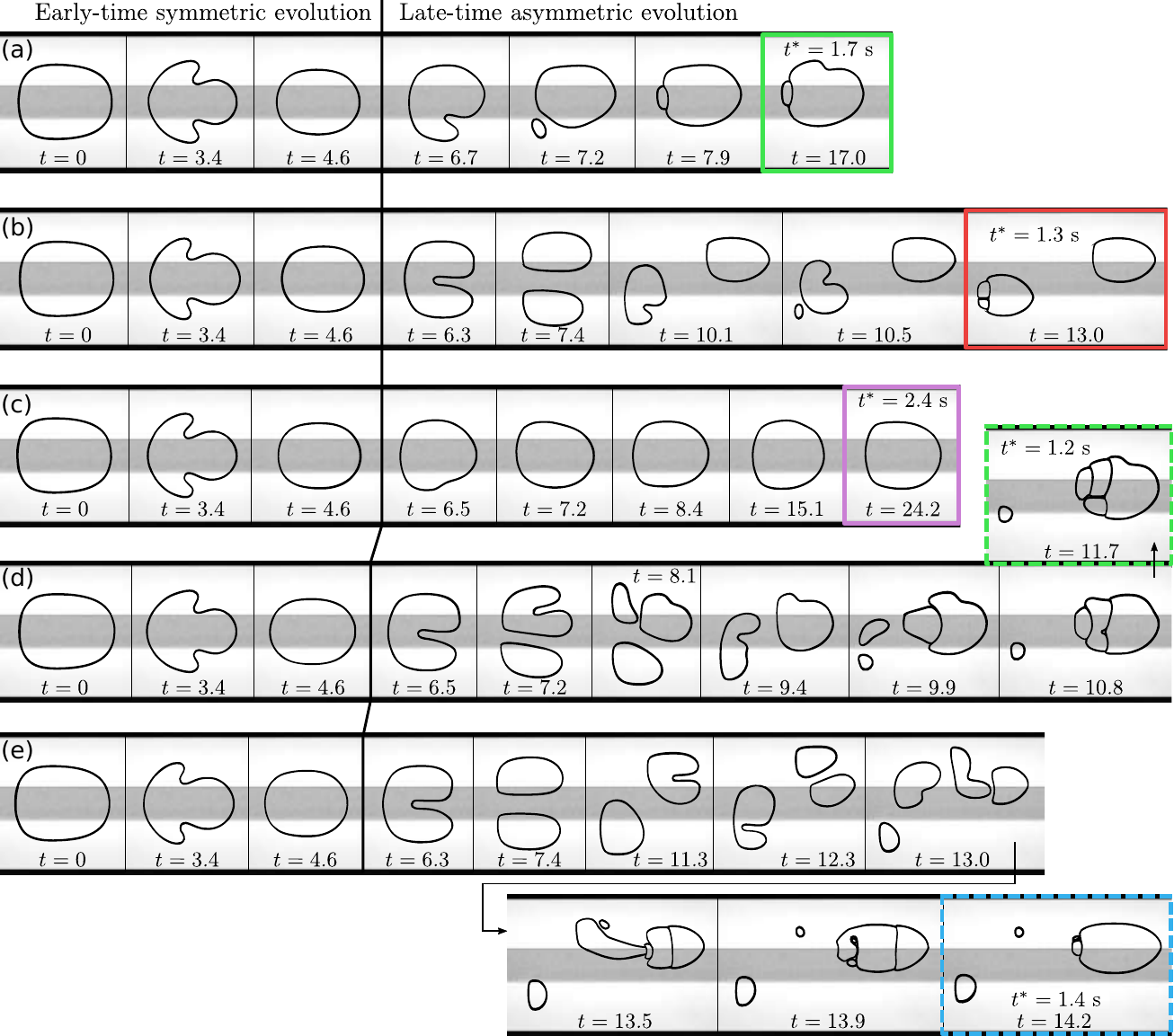}}
	\caption{Evolution of a bubble of initial width $d = 57.3$~\% propagating from rest at a flow rate $Q^* = 477$~mL/min ($Q = 0.18$) with $r_Q=0.54$. All experiments are performed for the same experimental parameters and lead to the formation of a transient approximately symmetric bubble ($t=4.6$) which, except in (c), undergoes tip-splitting and evolves towards different final outcomes. The time labelling is the same as in figure 4.}
	\label{fig:map_tristability}
\end{figure}

The time evolution becomes increasingly complex and unpredictable as the flow rate is increased, typically involving an increasing number of break up and aggregation events (hatched region in figure \ref{fig:mapexp}). This is illustrated in figure \ref{fig:map_tristability} where five successive experiments performed for parameter values corresponding to the yellow region in figure \ref{fig:mapexp} evolve towards five different final outcomes. At $t=0$, the relative difference in bubble area is within $0.5$~\% and the centroid positions along $x^*$ are within $1$~mm of each other. The time sequences in figures \ref{fig:map_tristability}(a, b) result in a single on-rail bubble ($t=17.0$) and to two off-rail bubbles ($t=13.0$) as tip-splitting occurs off-centre and on the centreline, respectively, similarly to figure \ref{fig:bug_break}. The flow rate is large enough for the on-rail, asymmetric bubble obtained in figure \ref{fig:map_tristability}(a) to oscillate. Figures \ref{fig:map_tristability}(d, e) provide examples of more complex time evolutions after centred tip-splitting, involving several break up and aggregation events. In figure \ref{fig:map_tristability}(d), an on-rail compound bubble (asymmetric and oscillating) is recovered followed by a tiny slowly propagating bubble, while in figure \ref{fig:map_tristability}(e), a large off-rail bubble is followed by two much smaller bubbles on either side of the rail, which will ultimately be sorted by size due to their differing velocities. A recurrent event during these complex evolutions is the rapid stretching and break up of bubbles trailing behind a leading bubble, whose products may in turn recombine with the bubbles ahead depending on their sizes and positions (figure \ref{fig:map_tristability}(d), $t=9.4 - 10.8$). In contrast, in figure \ref{fig:map_tristability}(c),  tip-splitting does not occur following initial symmetric deformation and the bubble remains on-rail, wobbling in a non-periodic manner with its centroid remaining on average at $y_c = 0$ until the end of the channel. This scenario only occurs for $Q > 0.155 \pm 0.005$ and becomes increasingly likely the larger the flow rate. 

The lack of reproducibility at high flow rate (figures \ref{fig:bug_break} and \ref{fig:map_tristability}) appears to stem from the increased sensitivity to both initial conditions and perturbations in the channel such as defects on the rail (of the order of the tape roughness). In order to test the influence of rail defects, we varied the portion of rail travelled by the bubble by initiating their propagation from different initial positions along the rail at high flow rate ($Q>Q_{ts}$) while keeping all other experimental parameters constant. By doing so, the bubble was subjected to different patterns of defects while propagating. We found that different initial bubble positions along the rail usually resulted in different locations of the dimple formed shortly after the bubble finishes its symmetric evolution. However, the associated tip-splitting always occurred at the same time after flow initiation for a given flow rate regardless of the initial bubble position. In contrast, when repeating experiments from the same initial position, the bubble generally followed the same late-time evolution, with rare exceptions. This suggests that the location of the dimple leading to tip-splitting coincides with the location of the largest rail defect ahead of the bubble once it reaches the end of its symmetric evolution. This is reminiscent of the findings of \citet{tabeling1987experimental} in a rectangular Hele-Shaw channel (in the absence of a rail), where symmetric fingers propagating at sufficiently large flow rates developed dimples at fixed positions in the channel due to localised perturbations on the top and bottom glass plates. 

However, in our experiments, the same initial bubble position (within $< 1$~mm) could also result in different dimple locations and therefore lead to different long-term outcomes. These relatively rare cases are those previously shown in figures \ref{fig:bug_break} and \ref{fig:map_tristability}. Ten experiments performed under the experimental conditions of figure \ref{fig:map_tristability} led to two off-centred tip splittings, five centred tip-splittings and three cases without tip-splitting like in figure \ref{fig:map_tristability}(c). We suspect that tip-splitting does not occur if the bubble does not encounter a sufficiently large rail defect at the end of its symmetric evolution phase. In all cases, the bubble fate is determined by its exact shape at that time, which exhibits small variations between experiments stemming from small differences in initial bubble shape at $t=0$.


\section{Comparison between experiment and numerical model}
\label{sec:interpretations}

 Interpretation of the phenomena described in \S \ref{sec:observations} based on the dynamical systems concepts presented in the introduction requires knowledge of the underlying stable and unstable invariant objects. In this section we present numerical results from a previously validated depth-averaged model that can be used to identify the appropriate invariant objects of the system.

\subsection{Depth-averaged model for bubble propagation}
\label{sec:Depth-averaged model for bubble propagation}

 The depth-averaged model extends \citeauthor{mccleantension}'s \citeyearpar{mccleantension} approach to account for a non-uniform channel height. We have previously used the same depth-averaged model to study the propagation of a semi-infinite air finger \citep{thompson2014multiple, franco2016sensitivity}
and a single closed air bubble \citep{franco2017bubble, franco2018bubble, keeler2019influence}
within the constricted channel. Two extensions are required here: (i) the simulation of multiple air bubbles; and (ii) the inclusion of topological changes in the form of coalescence and break up.

We work in a frame moving with the centroid position of the whole collection of bubbles and non-dimensionalise the physical system as shown in figure~\ref{fig:setup}: scaling the $x^*$ and $y^*$ directions by $W^*/2$, the $z^*$ direction by $H^*$ and the velocity by $U_0^* = Q^*/(W^*H^*)$.
The non-dimensional variable channel height is approximated by the smoothed tanh profile 
\begin{equation}
b(y) = 1 - \frac{1}{2}h\left[\tanh(s(y + w)) - \tanh(s(y - w))\right],
\end{equation}
where $h = h^*/H^*$ and $w = w^*/W^*$ are the non-dimensional height and width of the rail respectively and where $s$ sets the sharpness of the sides of the rail. We use the experimental values $h=0.024$ and $w = 0.25$, and choose $s=40$.

We describe the flow of the viscous fluid in terms of a depth-averaged velocity, assuming Stokes flow  (i.e. $\rho U_0^*W^*/(\mu\alpha^2) \ll 1$) and a large aspect ratio (i.e. $\alpha = W^*/H^* \gg 1$). Within the fluid domain, the 2D depth-averaged velocity $\mathbf{u}$ satisfies
\begin{equation*}
\mathbf{u} = -b^2(y) \nabla p, \quad \nabla \cdot (b(y)\mathbf{u})=0
\end{equation*}
where the second equation is a statement of mass conservation.
At the channel side walls, we enforce no-penetration conditions. The pressure is fixed to zero at the inflow, and a non-zero constant at the outflow to ensure the dimensionless volume flux is 2.

For both kinematic and dynamic conditions at the interface of each bubble, we assume that the air bubble fills the full height of the channel $b(y)$. Mass conservation then leads to the kinematic boundary condition
\begin{equation}
\frac{\partial\,\textbf{R}}{\partial\,t}\cdot \textbf{n} + \textbf{U}\cdot\textbf{n} + b^2(y)\nabla p\cdot \textbf{n} = 0.
\label{kinematic}
\end{equation}
at each bubble interface, where $\mathbf{R}$ is the 2D position of a point on the bubble boundary, $\mathbf{n}$ an outward normal unit vector,
and $\mathbf{U} = (U_b(t),0)$ with $U_b(t)$ the dynamically varying frame speed. 
We allow for multiple incompressible bubbles, each of different internal pressures. Hence the dynamic condition at the boundary of bubble $i$ is
\begin{equation}
p_i - p = \frac{1}{3\alpha Q}\left(\frac{1}{b(y)}+\frac{\kappa}{\alpha} \right),
\label{dynamic}
\end{equation}
where $p_i$ is the internal pressure in bubble $i$, $p$ the fluid pressure, $\kappa$ is the lateral curvature of the bubble in $(x,y)$ plane, and the transverse curvature contribution gives rise to the $1/b(y)$ term.

The pressure in each bubble, $p_i(t)$ is not known \textit{a priori}, but instead is determined so that the dimensionless volume $V_i$ of each bubble remains constant, where $V_i$ is calculated assuming that the bubble occupies the full height of the channel:
\begin{equation}
V_i = \int_{\Gamma_i} b(y)\, \mathrm{d}x\, \mathrm{d}y= -\int_{\partial \Gamma_i} x b(y) \,\mathrm{d}y,
\quad
A_i = \int_{\Gamma_i} \, \mathrm{d}x\, \mathrm{d}y.
\label{volume}
\end{equation}
Here $\Gamma_i$ is the interior of bubble $i$ and $\partial \Gamma_i$ its bounding curve. In the model, the dimensionless area and volume of the bubble are almost identical because we assume that the bubble fills the available height, which differs from 1 by at most $2.6\%$. As discussed in \S \ref{sec:Effect of the flow on the projected area of the bubble}, the projected area of the bubble $A_{i}$ varies with flow rate in the experiments, but it remains approximately constant at a given flow rate.

We discretise the system in space using a finite-element method, implemented in the open source \texttt{oomph-lib} package \citep{heil2006oomph}. The spatial discretisation uses a piecewise quadratic, triangular mesh, fitted to the bubble boundary. Mesh deformation is handled by treating the nodes of the mesh as embedded in a fictitious elastic solid. Time-stepping uses BDF1 or BDF2 (backwards difference formulae), with a new triangular mesh generated at least every 5 time steps. The channel length is typically taken as six times its width, to allow for interactions between bubbles of moderate separation.

The experimental results presented in \S \ref{sec:observations}  show frequent changes in bubble topology and we must therefore allow for both coalescence and break up in our numerical simulations.
To the best of our knowledge, there is no numerical method for solving our model equations in which bubble break up and coalescence do not require some intervention at the moment of topological change. In our implementation, the intervention involves detection of a topological event, reconnection of the lines defining the bubble interface, creating a new bulk mesh and restarting the simulation with a new set of volume constraints.
It is important to note that the only time derivatives in our model are those associated with motion of the interface. Hence after a change in topology, we need only specify the new configuration of the interface, and do not require an initial condition for the velocity of the fluid.

\begin{figure}
	\centerline{\includegraphics[scale=1]{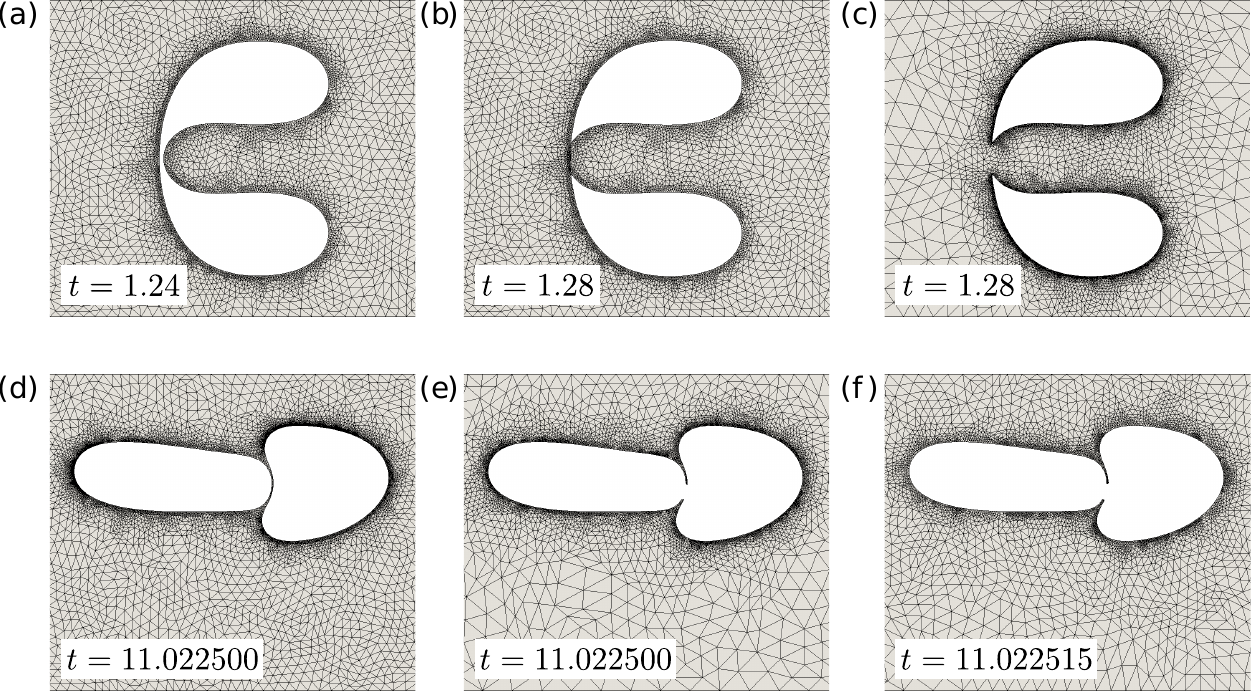}}
	\caption{Mesh interventions for break up and coalescence. (a) Bubble is simply connected before coalescence, $t=1.24$. (b) Self-intersection is detected, $t=1.28$. (c) Nodes in the intersected segment are removed, with the bubble reconnected via straight lines into two closed daughter bubbles, and the bulk is completely remeshed. (d) Two bubbles approach within the tolerance distance for coalescence, $t=11.022500$. (e) Nodes within the tolerance are removed, and the two bubble joined via straight lines. The bulk region is remeshed. (f) After several small time steps, the sharp corners from (e) have smoothed into recoiling tendrils, $t=11.022515$.}
    \label{fig:Topology}
\end{figure}

As the region inside the bubble is treated simply as a region of constant pressure and fixed volume, the need for break up is signalled by self-intersection of the bubble boundary, which  here happens in finite time (see figure \ref{fig:Topology}(a,~b)). After each timestep, we check for self-intersection. If intersection has occurred, we remove the intersected segments of the interface, and reconnect the two other segments with straight lines to create two simply connected daughter bubbles (figure \ref{fig:Topology}(c)). 
In contrast to break up, coalescence does not occur inevitably in finite time. Instead we must set some criteria that triggers coalescence, and we choose to base this on a simple minimum distance between the interfaces of two different bubbles. If bubbles approach below this minimum (figure \ref{fig:Topology}(d)), we remove the boundary nodes that are within the tolerance, and reconnect the remainder (figure \ref{fig:Topology}(e)).
Our topological `surgeries' for both coalescence and break up produce sharp corners on the bubble interface and are not volume conserving. To recover from the sharp corners, we take smaller time steps after a topological change, and remesh frequently. The interface typically regains a smooth shape very quickly (figure \ref{fig:Topology}(f)).  For coalescence, we set the target volume $V_i$ for the new bubble to be the sum of the two parent bubbles. 
For bubble break up, we make a small adjustment  to the target volumes for the two daughter bubbles (allocated in proportion to their measured volumes) in order to fully recover the volume of the parent bubble. 

We choose a dimensionless minimum distance for coalescence of $0.01$ which, although relatively large and significantly above the distance at which van der Waals forces become relevant, avoids the high computational cost of meshing very thin drainage regions. Indeed, once the drainage regions become thinner than the height of the channel our model is not expected to describe their behaviour accurately.

Comparison of bubble break up between model and experiments in our system indicates that break up is reasonably well approximated by our approach.
In contrast, the approach to coalescence is different between model and experiments. As described in \S \ref{sec:observations}, compound bubbles that can travel large distances before eventual coalescence is observed in the experiments. There is typically very little change in the propagation speed of the bubble and in the outline shape of the bubble front before and after coalescence. Therefore, we believe that the final state of the system is reasonably insensitive to details of the coalescence process, except perhaps in cases where bubble shapes are changing rapidly. The same general phenomena are observed in the simulations, but coalescence happens much more rapidly. 
We have confirmed that reducing the minimum distance for coalescence in the model to $0.001$  does not significantly affect the behaviour and merely increases the time taken for a compound bubble to coalesce.

In order to explore the time-dependent evolution of a single bubble we perform time simulations starting from a variety of initially elliptical bubble shapes. Specifically, we define the initial bubble shape as
\begin{equation}
x = l\cos(\theta),\qquad y = d\sin(\theta),\qquad \theta \in [0,2\upi].
\end{equation}
We note that this initial bubble shape is symmetric about the line $y=0$, but the triangular mesh generated in the fluid bulk is not perfectly symmetric.
All the initial value calculations presented here start from a single bubble with fixed initial area $A=\pi r^2 = \pi l d$ with $r=0.54$, with the bubble volume calculated from \eqref{volume}. We vary the flow rate $Q$ and the bubble width $d$ so that the phase diagram of the transient behaviour in the $(d,Q)$ plane can be compared with the experiment. 

Finally steady solutions can be calculated by setting the time derivatives to zero and the effects of the flow rate and volume can be mapped out using continuation and edge-tracking techniques \citep{Edge_track,gelfgat_bif}.

\subsection{Transient evolution and phase diagram}
\label{sec:Transient evolution}

\begin{figure}
	\centerline{\includegraphics[scale=1]{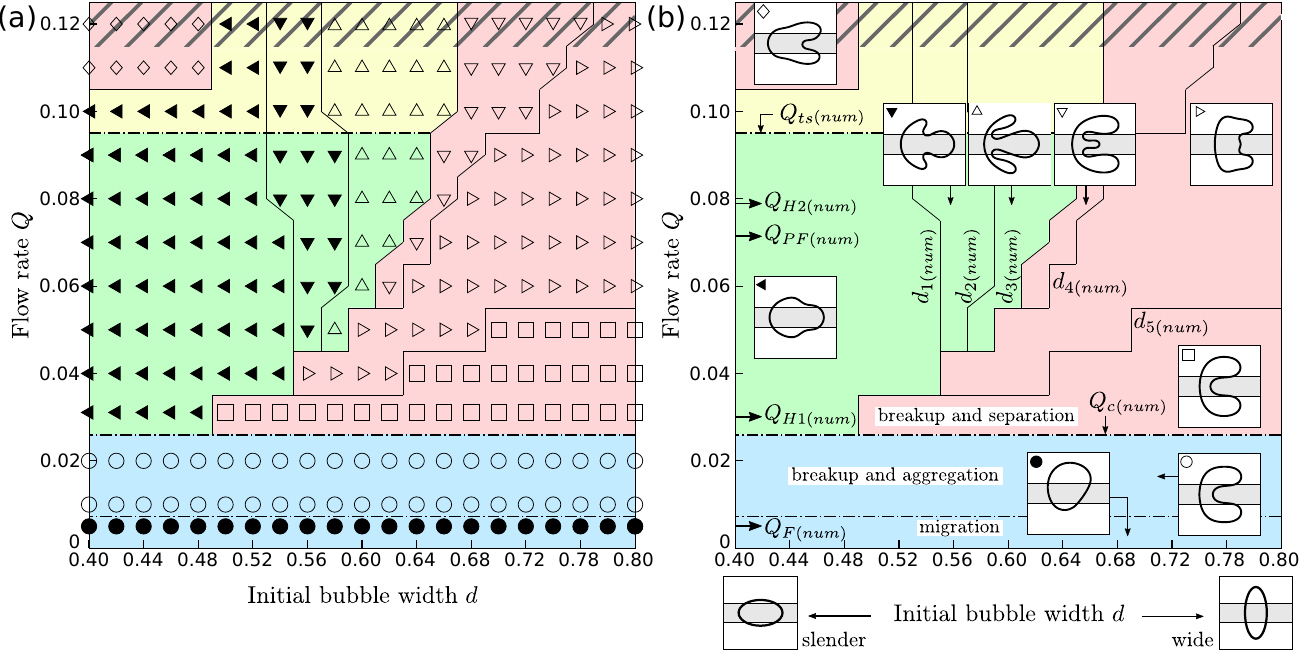}}
	\caption{Numerical phase diagram of the different time evolutions as a function of the non-dimensional initial bubble width $d$ and non-dimensional flow rate $Q$. The initial bubble shape is an ellipse of non-dimensional projected area $\upi ld$ and size $r = \sqrt{ld} = 0.54$. The layout, symbols, colours and notations are the same as in figure \ref{fig:mapexp} including the threshold values $Q_{c(num)}$ and $Q_{ts(num)}$. The symbols distinguish different transient evolutions shown with inset images of the associated initial deformations in (b). A ninth type of initial deformation is defined ($\diamond$) which involves a tip-split finger and occurs in the upper left corner of the diagram, for small values of $d$ which were not probed in the experiment. The values $Q_{F(num)}$, $Q_{H1(num)}$, $Q_{PF(num)}$ and $Q_{H2(num)}$ indicated by arrows in (b) are values calculated from a linear stability analysis indicating bifurcations in the solution structure of the single-bubble system; see figure \ref{fig:bifurcation_num}. The last three values mark transitions between different stable one-bubble invariant solutions. Specifically, the long term outcomes observed are off-rail bubbles for $Q< Q_{H1(num)}$, symmetric on-rail bubbles for $Q_{H1(num)} \le Q < Q_{PF(num)}$, asymmetric on-rail bubbles for $Q_{PF(num)} \le Q < Q_{H2(num)}$ and oscillatory on-rail bubbles for $Q \ge Q_{H2(num)}$.}
	\label{fig:map_num}
\end{figure}

The model described in \S \ref{sec:Depth-averaged model for bubble propagation} captures qualitatively all the bubble evolutions observed experimentally. A total of 273 simulations performed for $Q \le 0.12$ are organised in a numerical phase diagram in figure \ref{fig:map_num}. The phase diagram uses the same structure, symbols and colours as the experimental phase diagram shown in figure \ref{fig:mapexp}.

 The initial bubble deformations and long-term outcomes are qualitatively similar in both experimental and numerical phase diagrams, although the threshold flow rates do not coincide exactly. For example,
 the blue-red border separating one-bubble from two-bubble long-term outcomes occurs at a threshold flow rate $Q_{c(num)}=0.025 \pm 0.005$ (compared to the experimental value $Q_c=0.017 \pm 0.001$).
 The single-bubble invariant states arise in the same order as in the experiment as $Q$ is increased: off-rail bubbles ($Q < Q_{H1(num)}$); symmetric on-rail bubbles ($Q_{H1(num)} \le Q < Q_{PF(num)}$); asymmetric on-rail bubbles
($Q_{PF(num)} \le Q < Q_{H2(num)}$); and finally oscillatory on-rail bubbles ($Q \ge Q_{H2(num)}$); see \S \ref{sec:Bifurcation diagram} for a discussion of the associated bifurcation structure. 

\begin{figure}
	\centerline{\includegraphics[scale=1]{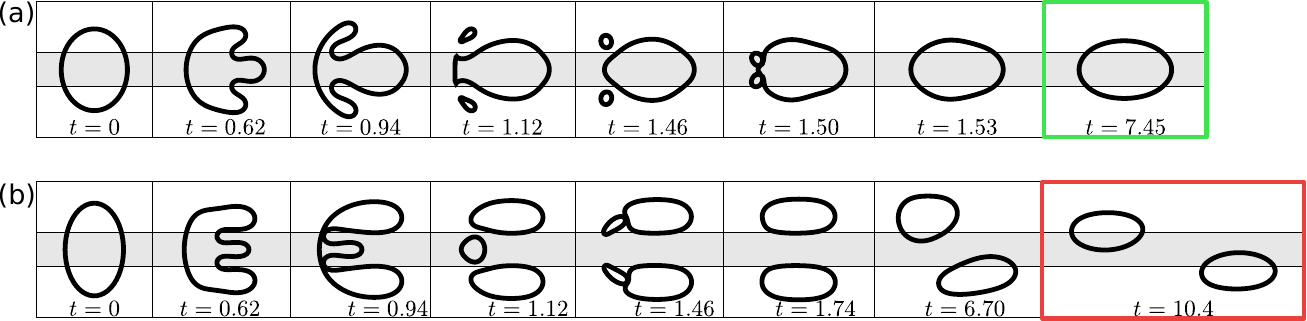}}
	\caption{Numerical time evolutions of an initially centred elliptical bubble with parameters $Q=0.07$ and $d=0.60$ (symbol: $\bigtriangleup$) (a) and $Q=0.10$ and $d=0.68$ (symbol: $\bigtriangledown$) (b); see figure \ref{fig:map_num}. The dimensionless time is given at the bottom of each snapshot. The bubble size is $r=0.54$.}
	\label{fig:num_snapshots}
\end{figure}

Two examples of numerical time evolutions are shown in figure \ref{fig:num_snapshots}(a, b), which are representative of regions in figure \ref{fig:map_num} indicated by symbols $\bigtriangleup$ and $\bigtriangledown$, respectively. These time sequences illustrate the key role of the relative size of side and middle tips in determining the long-term outcome when the front of the bubble initially deforms into three tips, consistent with experimental findings.

A distinction from the experimental phase diagram is the increased range of initial bubble widths $0.40 \le d \le 0.80$ in figure \ref{fig:map_num} compared to the experimental range in figure \ref{fig:mapexp}. For $d \le 0.48$ (not accessed experimentally), numerical simulations reveal only single-bubble long-term outcomes at moderate flow rates, as indicated by the blue-green border in figure \ref{fig:map_num}. At high flow rates within this range of initial bubble widths, we identify a ninth type of bubble evolution where the bubble breaks in two parts of unequal sizes following asymmetric tip-splitting ($\diamond$, red region in the top left corner of the phase diagram in figure \ref{fig:map_num}).

At flow rates $Q>Q_{ts(num)}$, where $Q_{ts(num)} = 0.095 \pm 0.005$ is close to the value $Q_{ts}=0.085 \pm 0.005$ measured experimentally, the late-time evolution of the bubble typically involves tip-splitting, like in the experimental evolutions of figures \ref{fig:bug_break} and \ref{fig:map_tristability}. In this yellow region of the phase diagram of figure \ref{fig:map_num}, small differences in the details of tip-splitting between two neighbouring data points may lead to either one or two-bubble final outcomes. More complex final outcomes occur in the hatched region as the complexity of the bubble evolution increases with flow rate. We hypothesise that the initial asymmetric tip-splitting observed in the ninth region of bubble evolution ($\diamond$) in figure \ref{fig:map_num} is analogous to the tip-splitting observed at late times in the yellow region but that it occurs sooner, superseding the initial symmetric deformation observed in the yellow region at early times.

The increasing sensitivity of the numerical model as $Q$ increases is reminiscent of the sensitivity to initial conditions and rail defects reported experimentally in \S \ref{sec:High flow rates}; and represented as a yellow and a hatched region in the phase diagram of figure \ref{fig:mapexp}. In the experiments variations occurred between repetitions at fixed parameter values due to unavoidable differences in the initial conditions.
In the numerical simulations the results are, of course, identical for the same initial conditions, but we find that smaller differences in initial conditions are required to provoke dynamically significant differences in the break up as the flow rate increases. In particular, modest changes in relative bubble volumes after break up can lead to quite different final outcomes, which may be a consequence of alterations in the associated invariant solution structure. 



Overall, the qualitative agreement between numerical simulations and experiments is remarkable given the simplicity of the mathematical model and approximation of the initial experimental conditions by ellipses.


\subsection{Single-bubble invariant modes of propagation}

\subsubsection{Bifurcation diagram}
\label{sec:Numerical Bifurcation diagram}

\begin{figure}
	\centerline{\includegraphics[scale=1]{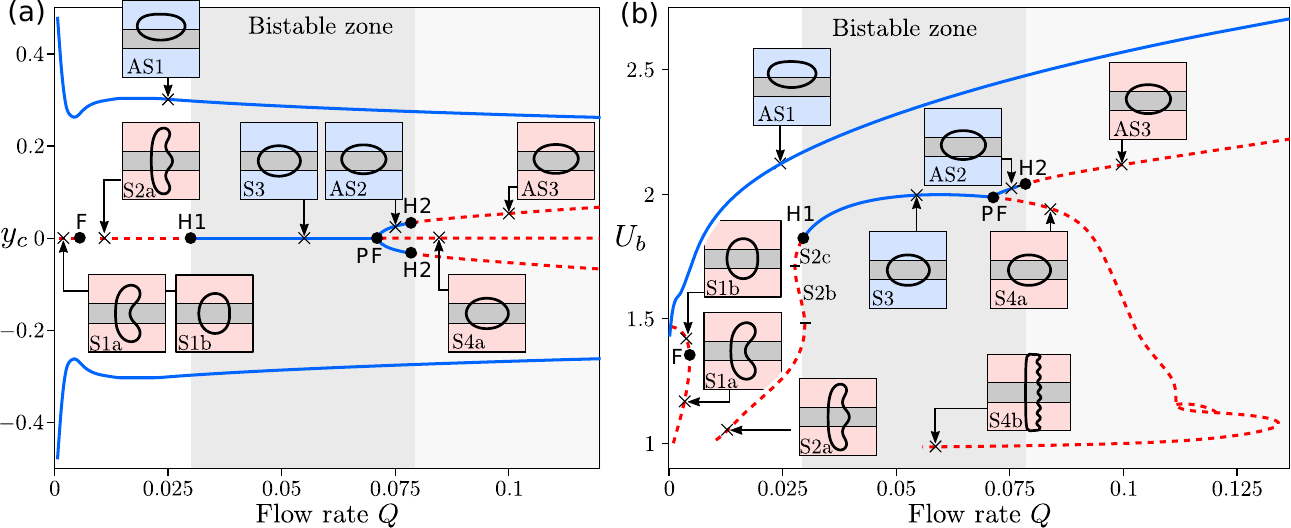}}
	\caption{Numerical bifurcation diagram showing steadily-propagating solutions for a single bubble. The non-dimensional bubble offset $y_c$ (a) and velocity $U_b$ (b) are plotted as a function of the non-dimensional flow rate $Q$ for a fixed bubble volume $V=0.54^2 \pi$ rather than a fixed projected area. Solid blue and dashed red lines indicate stable and unstable solutions, respectively. Each solution branch is illustrated by a snapshot corresponding to a flow rate indicated by a black cross, excepted of the S2b and S2c branches where snapshots are shown in figure \ref{fig:phase_plane_Q_0_029}. Solid circles indicate locations of bifurcation points. More details about the solution structure can be found in \citet{keeler2019influence}.}
	\label{fig:bifurcation_num}
\end{figure}

Steady solutions of the dynamical system are summarised in the numerical bifurcation diagram of figure \ref{fig:bifurcation_num} where the bubble centroid offset $y_c$ (a) and velocity $U_b$ (b) are plotted as a function of the flow rate $Q$ for a fixed bubble volume $V=0.54^2 \pi$. Stable and unstable solution branches are plotted with solid blue lines and dotted red lines, respectively. The stable solution branches are in agreement with the stable modes of propagation reported in the experimental bifurcation diagram of figure \ref{fig:bifurcation_exp}.

The numerical simulations indicate that $Q_{H1(num)}=0.0301$ marks a subcritical Hopf bifurcation, $Q_{PF(num)}=0.0714$ a supercritical pitchfork bifurcation and $Q_{H2(num)}=0.0789$ a supercritical Hopf bifurcation, as previously reported by \citet{keeler2019influence}. In the experiments, the first two bifurcations occur at values of $Q$ lower by factors of $1.9$ and $1.6$, while $Q_{H2}$ is larger than its numerical counterpart by a factor of $1.5$.

In addition, the numerical bifurcation diagram reveals the organisation of the unstable solution branches. We will discuss their influence on the bubble evolution in \S \ref{sec:single-bubble unstable solutions}, but we note the absence of any symmetric solution for $Q > 0.1345$, where the unstable symmetric branch S4 reaches a limit point.

\subsubsection{Single-bubble edge states}
\label{sec:single-bubble unstable solutions}

\begin{figure}
	\includegraphics[scale=1]{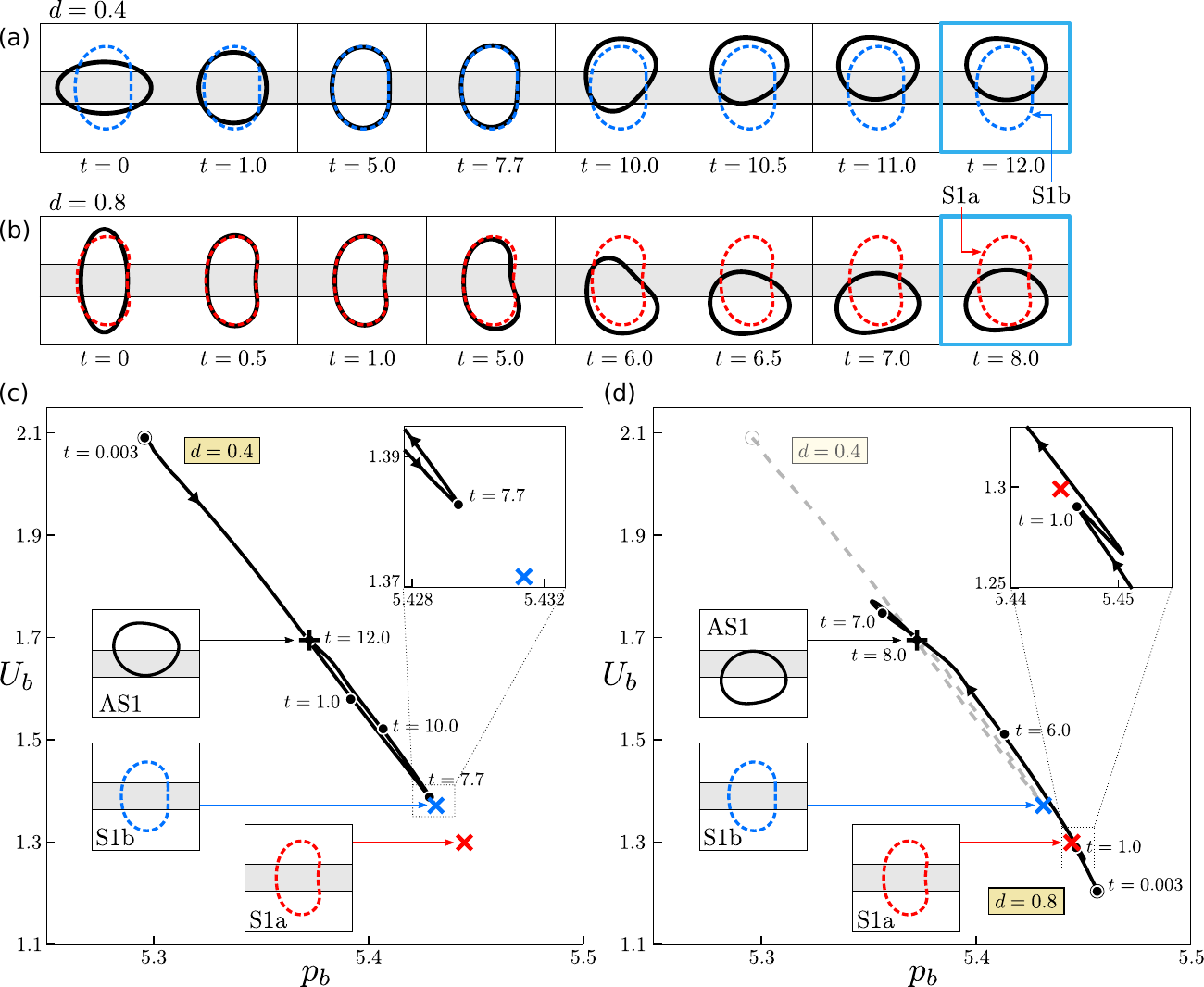}
	\caption{Numerical time evolution of initially elliptical bubbles of width $d=0.4$ (a) and $d=0.8$ (b) and size $r = 0.54$. The flow rate $Q=0.005$ is just below the value at which the S1a and S1b branches disappear through a limit point. The dashed bubble contours correspond to the S1b (a) and S1a (b) unstable steady states; see figure \ref{fig:bifurcation_num}. In both cases, the bubble eventually reaches the stable AS1 solution, although in (a) it converges to the `upper' asymmetric branch and in (b) to the `lower' asymmetric branch. Trajectories of the system are shown in (c) and (d) in a projection of the phase space $(p_b, U_b)$.  In (d), the dashed line reproduces the trajectory shown in (c). Some of the times indicated in the snapshots (a) and (b) are reported on the trajectories in (c) and (d), respectively. Coloured crosses correspond to the positions of the S1a and S1b steady states. Note that in (c) the trajectory appears to explore AS1 twice because the phase-space projection does not capture offset variations, but in fact the bubble only reaches AS1 at late times.}
	\label{fig:phase_plane_Q_0_005}
\end{figure}

The symmetric unstable solution branches associated with double-tipped (S1a and S1b) and triple-tipped (S2) bubbles, calculated numerically for the theoretical model and reported in figure \ref{fig:bifurcation_num}, are strikingly reminiscent of the early-time bubble shapes transiently observed in experiments; see figure \ref{fig:mapexp}. 
We also observe that the order in which the double and triple-tipped shapes appear in the numerical solution space is consistent with the order in which the corresponding shapes are observed experimentally as a function of the flow rate, as also reported by \citet{franco2018bubble}. 
We believe that broadly equivalent unstable steady states exist in the model and the experiment.  We now show that these unstable states have a dramatic impact on the transient dynamics of the system as a whole, with the richness of the unstable solution structure reflected in the transient dynamics.

In the model, the first pair of solution branches exists only for very low flow rates, merging at a limit point at $Q_F=0.005085$. The S1b and S1a branches have respectively one and two unstable eigenvalues and only the shape associated with the S1a branch features two pronounced tips.
In the experiments, shapes similar to the S1b and S1a solutions are transiently explored at $Q=0.005$ in figures \ref{fig:map_vertical}(a) ($t=2.4$) and  \ref{fig:map_vertical_wide}(a) ($t=2.3$), respectively, before the bubble eventually migrates towards the stable AS1 solution. Hence it appears that slender bubbles are initially attracted towards an equivalent of the S1b solution via its symmetric stable manifold and are finally repelled via its asymmetric unstable manifold, while wider bubbles are attracted to the symmetric S1a solution before being repelled asymmetrically.

\begin{figure}
	\includegraphics[scale=1]{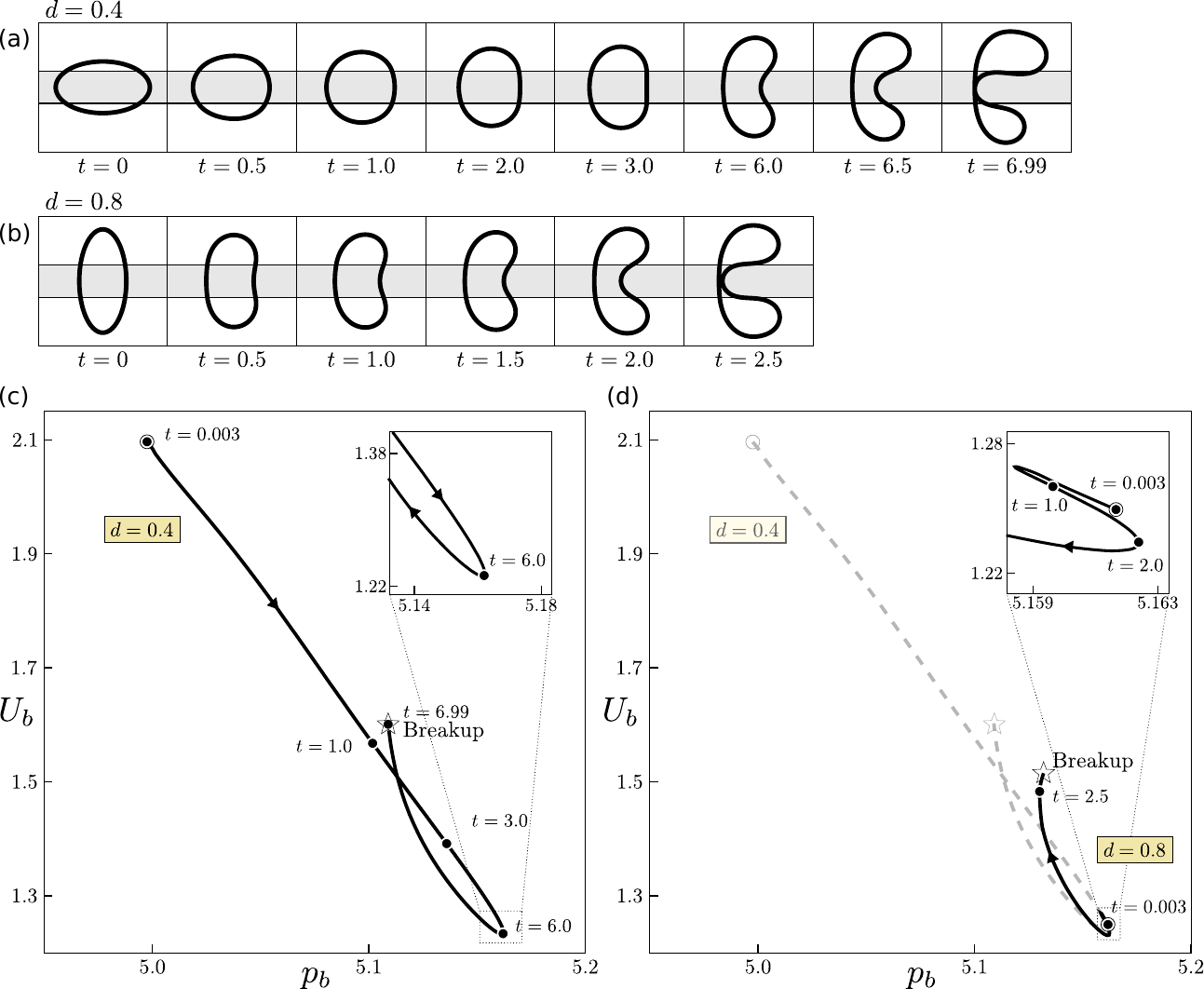}
	\caption{Numerical time evolution of initially elliptical bubbles of width $d=0.4$ (a) and $d=0.8$ (b) and size $r = 0.54$. The flow rate $Q=0.006$ is just above the value at which the limit point $F$ occurs so that the S1 branch does not exist; see figure \ref{fig:bifurcation_num}. (c, d) Trajectories are shown in a ($p_b$,$U_b$) projection of the phase space with a layout similar to that of figure \ref{fig:phase_plane_Q_0_005}.}
	\label{fig:phase_plane_Q_0_006}
\end{figure}

We can test this claim by studying a numerical time simulation at a low flow rate ($Q=0.005$, with the limit point at $Q_F = 0.005085$). Figure \ref{fig:phase_plane_Q_0_005} shows the evolution of a single bubble from initial elliptical shapes of width $d = 0.4$ (a) and $0.8$ (b). The bubble contours corresponding to the S1b and S1a unstable solutions are shown with coloured dashed lines in figures \ref{fig:phase_plane_Q_0_005}(a) and (b), respectively. 
At $Q=0.005$ the S1a solution has two unstable eigenvalues $\lambda= 1.43560$ and $\lambda=0.27542$ (the time dependence is proportional to $\exp(\lambda t)$), whilst the S1b solution has a single unstable eigenvalue $\lambda=1.09033$. 

In figure \ref{fig:phase_plane_Q_0_005}(c, d) we visualise the time evolution of the bubbles shown in figure \ref{fig:phase_plane_Q_0_005}(a, b) as trajectories projected into a two-dimensional phase plane spanned by bubble pressure $p_b$ and bubble velocity $U_b$, where the steady solutions S1a and S1b are indicated with coloured crosses. 
Figure \ref{fig:phase_plane_Q_0_005}(c) reveals that initially slender bubbles ($d=0.4$) visit the symmetric stable manifold of the least unstable (S1b) solution before leaving the vicinity of S1b via its asymmetric unstable manifold and eventually settling on the stable asymmetric AS1 solution (see inset in (c)).
Initially wider bubbles ($d=0.8$) in figure \ref{fig:phase_plane_Q_0_005}(d) are instead attracted to the more unstable (S1a) solution resulting in a more visible two-tipped deformation. After leaving the vicinity of the S1a solution the system settles again on an AS1 solution. As shown in figure \ref{fig:phase_plane_Q_0_005}(a, b), the shape of the evolving bubble and that of the associated unstable steady state are virtually indistinguishable at times $t = 5.0$ and $t = 1.0$, respectively. Note, however, that the chosen phase plane projection does not capture a complete picture because it does not, for example, convey information about the bubble offset.

\begin{figure}
	\includegraphics[scale=1]{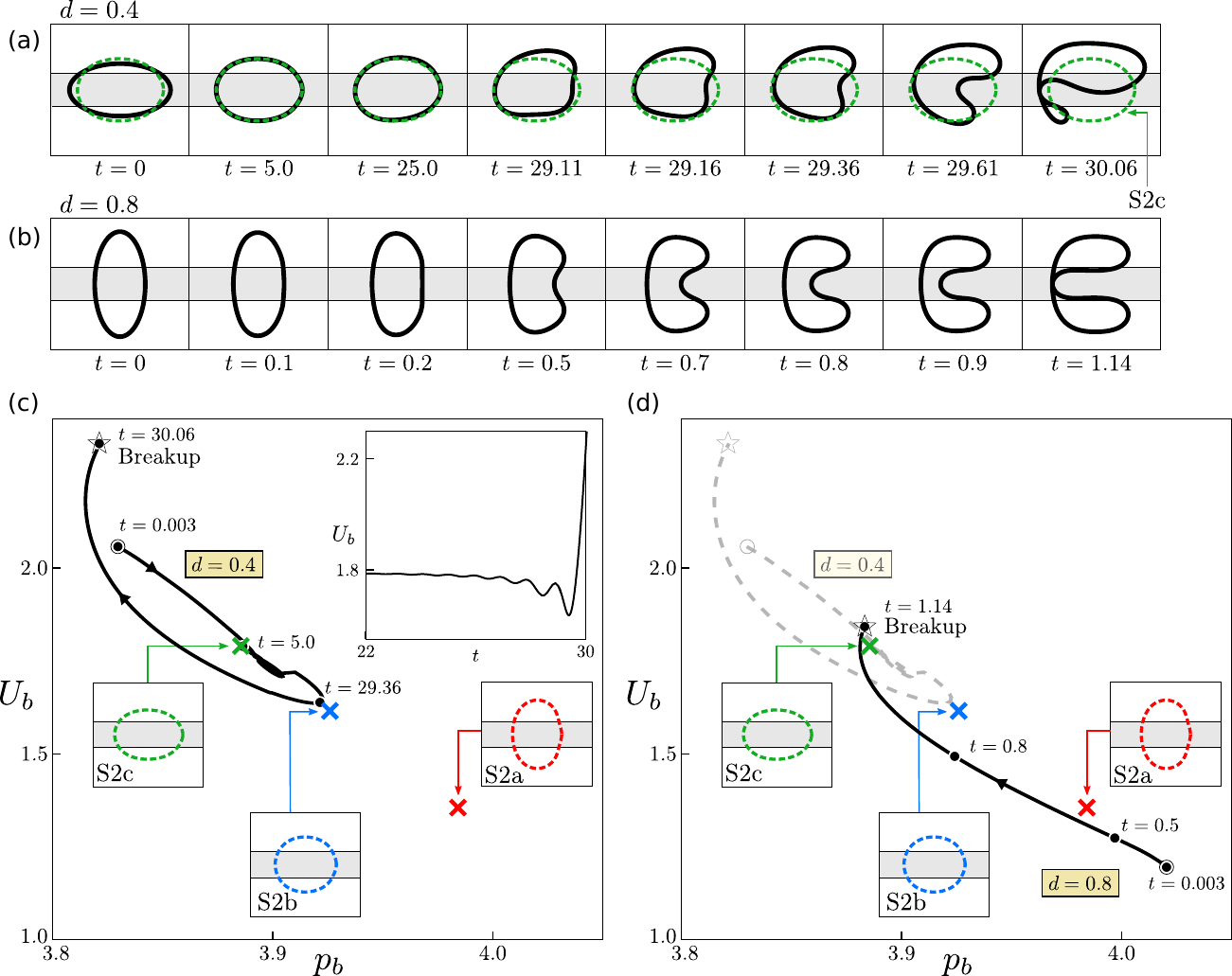}
	\caption{Numerical time evolution of initially elliptical bubbles of width $d=0.4$ (a) and $d=0.8$ (b) and size $r = 0.54$ for $Q=0.029$. The dashed contours in (a) indicate the S1c unstable steady state; see figure \ref{fig:bifurcation_num}. (c, d) Trajectories are shown in a ($p_b$,$U_b$) projection of the phase space with a layout similar to that of figure \ref{fig:phase_plane_Q_0_005}. Coloured crosses in (c) and (d) correspond to the positions of the S2a, S2b and S2c steady states; see figure \ref{fig:bifurcation_num}. In both cases, the bubble eventually breaks up, but the trajectories are very different. The initially slender bubble lingers for a long time near the S2c solution, which is weakly unstable with oscillatory eigenmodes, before eventually breaking up asymmetrically. The initially wide bubble experiences a much more straightforward evolution towards break up into two equal bubbles.}
	\label{fig:phase_plane_Q_0_029}
\end{figure}

The highly deformed double-tipped shape is also transiently observed in the numerical calculations for larger values of $Q$, for which the S1 branch does not exist. Figure \ref{fig:map_num} shows that for $0.010 \le Q \le 0.025$ the two-tipped shape is seen for all bubble widths and also for $Q \ge 0.025$ when the bubble is sufficiently wide. 
Although the S1 branch does not exist in this range, the trajectories in phase space are expected to vary smoothly beyond the limit point at $Q_F$. Hence it is plausible that the system
will trace a trajectory through phase-space that results in two-tipped bubble shapes even if the steady solution is not present. Figure~\ref{fig:phase_plane_Q_0_006} shows the trajectories of slender and wide bubbles at $Q=0.006$ where the S1 branch does not exist. For both cases the trajectory initially follows a qualitatively similar path through phase space compared to when $Q=0.005$. However the late evolution is different in that rather than reaching the stable AS1 state directly the bubble breaks up. 
It appears the unstable S1 branch plays a vital role in enabling a single bubble to approach the AS1 state as its absence means that the bubble will break up. Therefore the value of $Q_F$, the highest flow rate for which the unstable branch exists, marks the threshold between the solid and hollow circular markers in figure~\ref{fig:map_num}.

The same approach can be used to examine the influence of the unstable S2 solution on the dynamics of the system.
Figure~\ref{fig:phase_plane_Q_0_029} shows trajectories for a slender and wide bubble when $Q=0.029$. At this flow rate, in addition to the stable AS1 solution, there are three unstable steady states on the S2 branch. The most unstable S2a solution has four unstable eigenvalues $\lambda =1.29566$, $2.94015\pm1.48472i$, 4.15816, the S2b solution has three unstable eigenvalues $\lambda=1.44914\pm2.51394i$, 1.85719 and the S2c solution has a complex conjugate pair of unstable eigenvalues $\lambda =0.19702\pm 3.54153i$. For an initially slender bubble ($d=0.4$) the system evolves towards the S2c solution where it stays in the stable manifold for a long time, see figure~\ref{fig:phase_plane_Q_0_029}(a, c). Due to the complex eigenvalues the trajectory oscillates in the vicinity of S2c (see the inset time signal of $U_b$ in (c)) before eventually breaking up in a highly asymmetric manner that resembles a thwarted migration towards the stable AS1 solution. In contrast the dynamics for an initially wide bubble ($d=0.8$) are significantly quicker resulting in a symmetric break up event.
These results demonstrate not only that the unstable states are transiently observed during time evolution, as claimed, but also that their presence influences the dynamics of single bubbles.



\subsection{Two-bubble edge states}
\label{sec:Interaction between two bubbles of comparable sizes}

\begin{figure}
	\centerline{\includegraphics[scale=1]{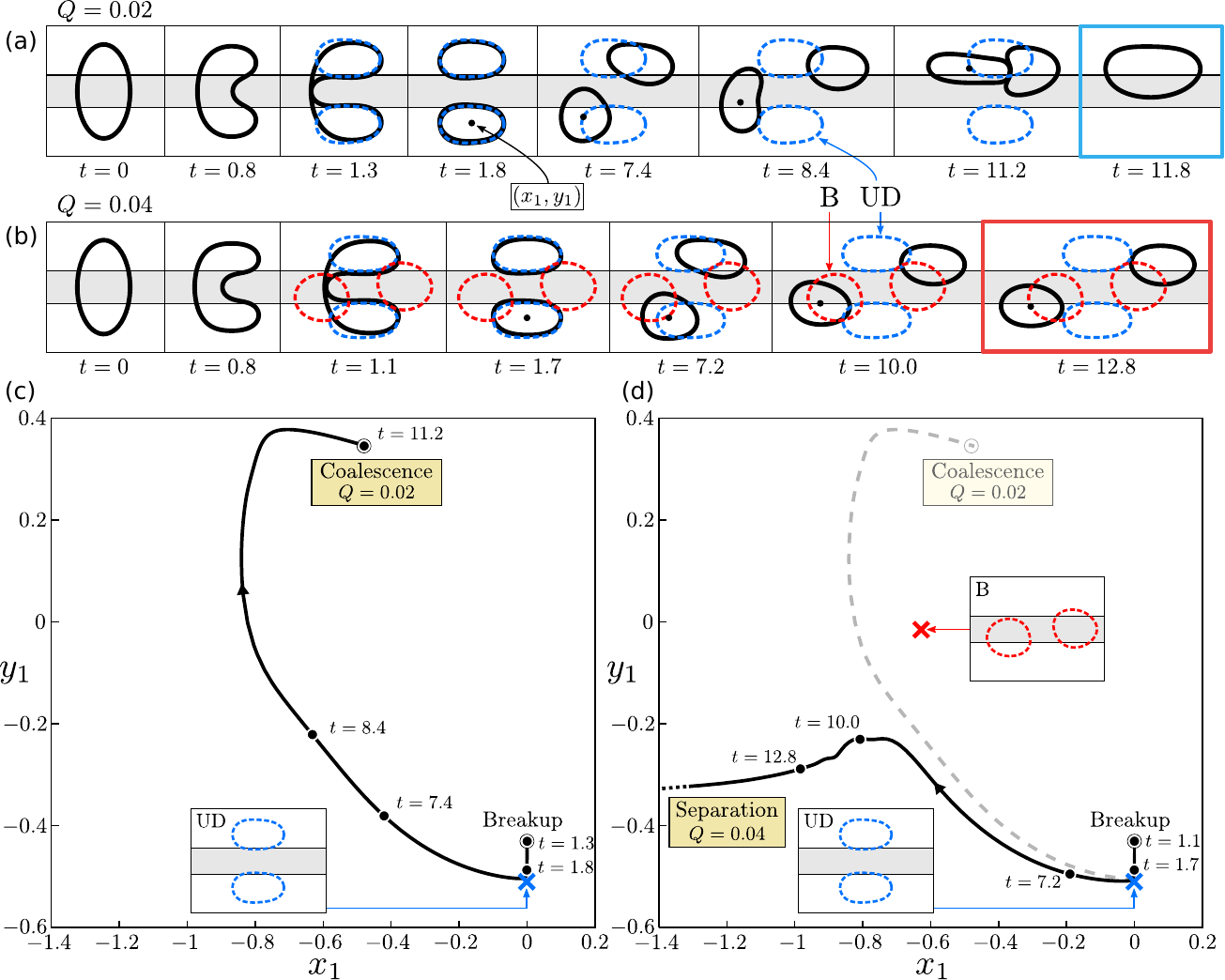}}
	\caption{Numerical time evolutions at flow rates $Q=0.02$ (a) and $Q=0.04$ (b) for an initially elliptical bubble of width $d=0.72$ and size $r=0.54$. The dimensionless time is given at the bottom of each snapshot. The two-bubble coloured dashed shapes correspond to unstable steady states of the two-bubble system, labelled `UD' for `up-down' (blue) and `B' for `barrier' (red). According to our depth-averaged model, the B solution emerges for flow rates greater than $0.029 \pm 0.001$ in our model. The position $(x_1,y_1)$ of the centroid of the trailing bubble identified by a point. The associated trajectories of the two-bubble system in the $(x_1,y_1)$ phase space projection are represented by the solid lines in (c) ($Q=0.02$) and (d) ($Q=0.04$). In (d), the dashed line is a reproduction of the trajectory in (c). Times corresponding to the snapshots of (a) and (b) are reported on the trajectories (c) and (d) respectively. Coloured crosses in (c) and (d) correspond to the values of $(x_1,y_1)$ associated with the `UD' and `B' unstable steady states.}
	\label{fig:2bubble_edge_states}
\end{figure}

In both the experiments and numerical simulations at relatively low flow rates, an initially wide bubble splits in two and a critical flow rate $Q_{c}$ determines whether the two bubbles coalesce or separate indefinitely, see figures \ref{fig:mapexp} and \ref{fig:map_num}. We now provide evidence that after bubble break up unstable two-bubble steady states are so-called edge states responsible for the organisation of these dynamical outcomes of the system.

Numerical simulations of the time evolution of an initially elliptical bubble of width $d=0.72$ at $Q=0.02$ are shown in figure \ref{fig:2bubble_edge_states}(a) demonstrating that the bubble splits into two bubbles which later coalesce. Simulations of the evolution of the same bubble at a higher flow rate, $Q=0.04$, are shown in figure \ref{fig:2bubble_edge_states}(b) and in this case the two bubbles do not coalesce. The computed time evolutions are qualitatively similar to the experimental evolutions shown in figures \ref{fig:map_vertical_wide}(b) and (c) respectively. Steady simulations reveal that there is an unstable steady `up-down' (UD) two-bubble solution at $Q=0.02$, shown as blue dashed lines in figure \ref{fig:2bubble_edge_states}(a). At $Q=0.04$, in addition to the UD solution there is another `barrier' (B) unstable steady solution, shown as red dashed lines in figure \ref{fig:2bubble_edge_states}(b). 

In order to demonstrate how the dynamics is related to the identified steady states of the two-bubble system, figures \ref{fig:2bubble_edge_states}(c) and (d) show the time evolutions in a projection of the phase space of the system created by plotting the position $(x_1,y_1)$ of the centroid of the trailing bubble when two bubbles are present. Note that the projected phase space is not the same as that in \S \ref{sec:single-bubble unstable solutions}. The steady two-bubble solutions are also marked on the same projection of the phase space.
The two bubbles are formed at the point labelled `break up' and the subsequent evolution is shown by the solid black line. In both cases, the two bubbles initially propagate side-by-side as the system is attracted towards the UD steady state, which has only a single unstable eigenvalue and is therefore strongly attracting. As seen in figures \ref{fig:2bubble_edge_states}(a) and (b) the bubble shapes are virtually indistinguishable from those of the UD steady state at $t=1.8$ and $t=1.7$ respectively. The system then moves away from the unstable steady state in the direction associated with the single unstable eigenvalue: in the projection of the phase space, the unstable steady state is analogous to a two-dimensional saddle point.

At $Q=0.02$, the lower bubble moves up and over the rail and two bubbles aggregate and coalesce. In terms of movement through the phase space, the system is attracted to a compound bubble that is analogous to a stable single-bubble state. Examples of such compound bubbles are found throughout \S \ref{sec:observations}.
At $Q=0.04$ the presence of the second unstable steady state and its associated unstable manifold prevents the system from reaching a coalescence point by deflecting the trajectory. Although the lower bubble starts to move over the rail it remains on the lower side of the rail because the B state is sufficiently repulsive. Unable to reach the coalescence point, the two bubbles continue to separate because they move at different speeds. The B steady state arises through a saddle-node bifurcation and only exists for Q greater than a threshold flow rate $0.029 \pm 0.001$ which can hence be identified with the value $Q_{c(num)}$ roughly estimated in figure \ref{fig:map_num}.

These calculations provide compelling evidence that the unstable two-bubble steady states are edge states whose stable and unstable manifolds partition the dynamical outcomes of the system. The two-bubble system is more complicated than the single bubble case, in part because the volume ratio between the two bubbles is an additional parameter. For example, if the bubble is not initially symmetric, then break up may result in two bubbles of significantly different sizes for which the impact of the edge states described here may differ from figure \ref{fig:2bubble_edge_states}. For all size ratios we have examined, we find that every single-bubble steady state has an equivalent two-bubble steady state with near identical flow rate--propagation speed characteristics and hence these solution branches are qualitatively unchanged. There are, however, additional two-bubble steady states that do not have single-bubble equivalents and a thorough investigation and discussion of two-bubble steady states is provided in the companion paper \citep{Keeler2020}.

\section{Conclusion and perspectives}
\label{sec:conclusion}

We have studied the dynamics of a bubble evolving from a prescribed initial condition as it propagates through a Hele-Shaw channel driven by the steady motion of a suspending viscous fluid. We have focused on a channel with a depth perturbation in the form of a centred rail, because it supports multiple invariant modes of bubble propagation which lay the foundation for complex dynamics. These invariant modes can be distinguished using only the interfacial morphology because the nonlinearity in this system stems from the presence of the air-oil interface. We observe a rich array of organised transient behaviour, which may involve repeated bubble break up and coalescence events, before a long-term outcome is reached in the form of either a single-bubble invariant mode of propagation or two (or more) steadily separating bubbles (sorted in order of decreasing size and velocity). A highly unusual feature of this system is that the underlying invariant-solution structure evolves dynamically, because changes in bubble topology causes the number of bubbles to vary.

A depth-averaged model of bubble propagation captures the majority of the experimental dynamics and features similar single-bubble, stable invariant states as observed in the experiment. We therefore conjecture that broadly similar unstable invariant states occur in the experiment as in the model and we find that the bubble transiently adopts the shape of unstable solutions during its evolution. Moreover, we show that multiple unstable invariant states play a key role in organising the transient dynamics as edge states, by partitioning the dynamical outcomes of the bubble evolution via their stable and unstable manifolds. The multiplicity of edge states in turn underpins the subtle dependence of the dynamics on the width of the initial bubble. Initial conditions determine the specific paths of exploration between these edge states and thereby the time scales of transient evolution which can vary significantly.

We identify the specific two-bubble edge state that controls the fate of the bubble once it has broken up: in its absence the smaller trailing bubble migrates across the rail to coalesce with the leading bubble, while in its presence the products of the break up remain on opposite sides of the rail and separate indefinitely. The complexity of the two-bubble system is such that we defer its detailed analysis to a forthcoming paper \citep{Keeler2020}. A particularly striking feature to be discussed is that all single-bubble invariant states have equivalents in the two-bubble system. However, additional multiple-bubble states mean that the complexity of the dynamics increases with the number of bubbles.

For moderate flow rates, the evolution of the bubble is fully reproducible in the experiment. However, as the flow rate increases, the experiments exhibit enhanced sensitivity to both initial bubble condition and unavoidable rail perturbations. Hence, multiple repetitions of the same experiment lead to widely different long-term outcomes, e.g., multiple separating bubbles versus a single-bubble invariant mode of propagation.  An analogous sensitivity is also found in numerical time simulations for comparably high values of the flow rate. This sensitivity makes the long-term behaviour of the system practically unpredictable.

In both the experiments and simulations the evolution of the bubble at high flow rate follows a complex scenario of transient emulsification and coarsening through repeated cycles of break up and coalescence. The sensitivity arises from differences in the break up of the bubble, rather than direct sensitivity to initial conditions per se, and slight differences in the relative volumes of the daughter bubbles that are created can lead to very different subsequent dynamics. We conjecture that the increased complexity of the dynamics is associated with a growing number of invariant states: the number increases with increases in the flow rate and also with increases in the number of bubbles.
An increase in flow rate also promotes break up into an increasing number of bubbles with a broad size distribution which means that the number of potential invariant states can increase very rapidly. The large number of invariant states may entangle into a complex edge that need not persist for all time, but underlies long transients of disordered dynamics for sufficiently large flow rates. Whether the system can eventually exhibit self-sustaining disordered dynamics, equivalent to a fully-developed turbulent state, remains an open question.

\vspace{2em}
\noindent \textbf{Acknowledgements}

This work was funded by EPSRC grant EP/P026044/1. We thank Martin Quinn for his technical support of the experimental facility.

\vspace{2em}
\noindent \textbf{Declaration of interests}

The authors report no conflict of interest.

\FloatBarrier
\appendix 
\section{Wetting films}
\label{sec:Wetting films}

\begin{figure}
	\centerline{\includegraphics[scale=1]{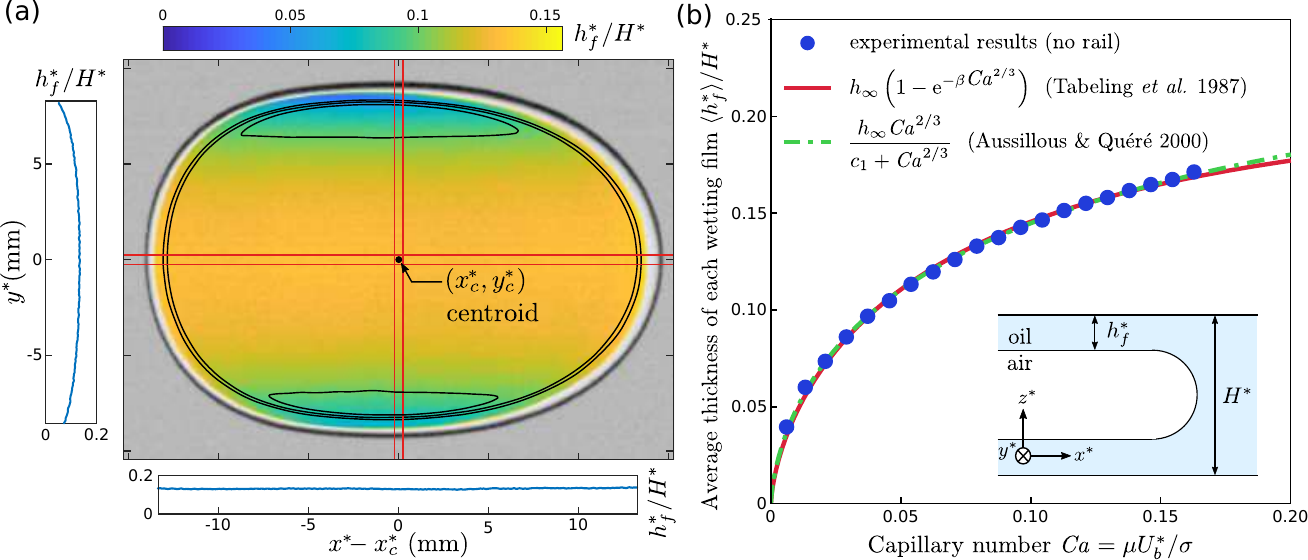}}
	\caption{(a) Spatial distribution of the wetting film thickness $h_f^*$ relative to the channel depth $H^*$ for a bubble of size $r_Q = 0.59$ propagating steadily and symmetrically about the channel centreline in the absence of a rail at a flow rate $Q^* = 100$~mL/min ($Q = 0.038$, $\Capi = 0.088$). The profiles along $x^*$ and $y^*$ are obtained by averaging over the horizontal and vertical bands indicated with red lines, which are centred on the centroid of the bubble ($x_c^*$,$y_c^*$). (b) Relative film thickness $\langle h_f^* \rangle / H^*$ averaged over the bubble area as a function of the capillary number $\Capi$. Two functions of $\Capi$ are fitted to the data: $\langle h_f^* \rangle / H^* = h_{\infty} (1-\exp{(-\beta \Capi^{2/3}}))$ \citep{tabeling1987experimental}, where the values of the fitting parameters are $h_{\infty} = 0.21$, $\beta = 5.6$ and $\langle h_f^* \rangle / H^* = h_{\infty} \Capi^{2/3} / (c_1 + \Capi^{2/3})$ \citep{Aussillous2000}, where $h_{\infty} = 0.31$ and $c_1 = 0.24$. Inset: Schematic view of the bubble tip in the $x^*$-$z^*$ plane.}
	\label{fig:wetting_films}
\end{figure}

 In order to estimate the thickness of the wetting films, we performed experiments using dyed silicone oil (APD fluorescent leak detection dye). The rail was removed for simplicity so that the bubble propagated steadily and symmetrically about the channel centreline. The total oil thickness $2h_f^*$ was estimated by measuring the intensity of the light transmitted through the channel from the LED light box to the top-view camera moving with the propagating bubble. Calibration of the light intensity was performed by measuring values for air and oil-filled channels, respectively. We also checked that the light intensity was proportional to the thickness of the attenuating medium by performing measurements through an oil-filled channel containing a transparent plastic wedge of fixed angle. In the following, we assume that the top and bottom oil films have the same thickness $h_f^*$.

A typical distribution of wetting film thickness over the projected area of the bubble is presented in figure \ref{fig:wetting_films}(a) for a bubble propagating at a flow rate $Q^* = 100$~mL/min. The bubble size during propagation is $r_Q = 0.59$. The film thickness distribution is symmetric about the channel centreline and uniform along $x^*$ for fixed values of $y^*$. Wetting films are thickest on the centreline ($y^*=0$) and thin as $|y^*|$ increases towards the edge of the bubble, which is consistent with the findings of \citet{tabeling1987experimental}.

We found that the film thickness $\langle h_f^* \rangle$ averaged over the projected area of the bubble does not depend significantly on the bubble size. The variation of $\langle h_f^* \rangle$ with the capillary number is shown in figure \ref{fig:wetting_films}(b). Two expressions have previously been used to capture this variation, both exhibiting the $\Capi^{2/3}$ behaviour predicted by \citet{bretherton1961motion} and \citet{park1984two} in the limit of vanishing $\Capi$. \citet{tabeling1987experimental} captured the film thickness of a semi-infinite bubble (finger) propagating in a rectangular Hele-Shaw channel with an empirical relation, $\langle h_f^* \rangle / H^* = h_{\infty} (1-\exp{(-\beta \Capi^{2/3}}))$, where the film thickness at large $\Capi$, $h_{\infty}$, and $\beta$ are fitting parameters. The scaling law $\langle h_f^* \rangle / H^* = h_{\infty} \Capi^{2/3} / (c_1 + \Capi^{2/3})$, where $h_{\infty}$ and $c_1$ are fitting parameters, was proposed by \citet{Aussillous2000} to characterise the thickness of the liquid film left behind an advancing finger in a capillary tube. Two-parameter fits of both expressions capture our experimental measurements, as shown in figure \ref{fig:wetting_films}(b).

\section{Influence of the bubble size}
\label{sec:Influence of the bubble size}

\begin{figure}
	\centerline{\includegraphics[scale=1]{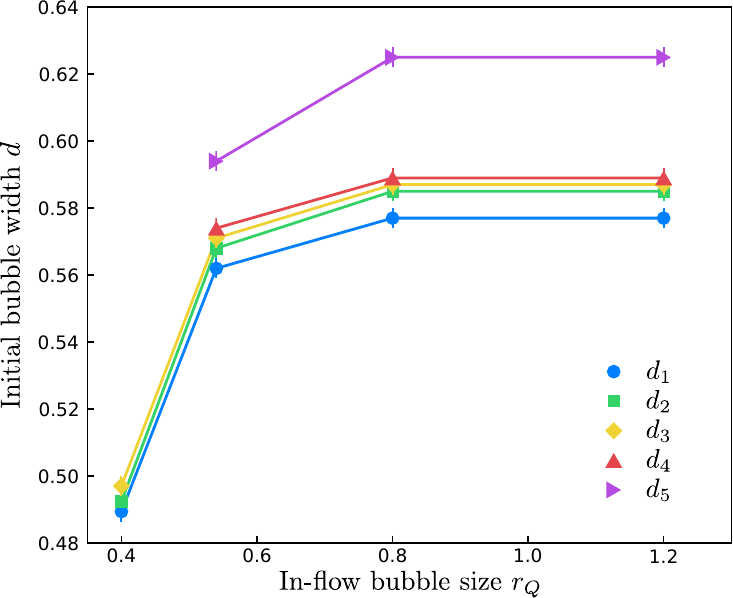}}
	\caption{Values of $d_i$ ($i=1,...5)$ at $Q=0.07$ (see figure \ref{fig:mapexp}) for bubbles of different in-flow sizes $r_Q = 0.4$, $0.54$, $0.8$ and $1.2$. For $r_Q = 0.4$, $d_4$ and $d_5$ could not be measured because they exceed the largest width $d_m \approx 0.45$ explored by the bubble during relaxation (see inset of figure \ref{fig:reshape}(c)).}
	\label{fig:influence_r}
\end{figure}

We investigated the influence of bubble size on the early-time symmetric evolution of the bubble. At $Q=0.07$, we observed early-time evolutions similar to that of figure \ref{fig:map_horizontal} ($r_Q=0.54$) when varying the bubble width $d$. The values of $d_i$ ($i=1,...5$) marking the boundaries between different transient evolutions (for $Q>0.025)$ are plotted in figure \ref{fig:influence_r} for different bubble sizes $r_Q$, revealing that they do not vary significantly for bubbles with $r_Q \ge 0.54$. This confirms that the initial width of the bubble $d$ influences the evolution of the bubble much more strongly than its initial length $l$. It also suggests that the time sequences of figures \ref{fig:map_vertical} and \ref{fig:map_vertical_wide}, where the time required to compress the bubble following the imposition of the flow increased with $Q$, exhibit behaviour representative of the system that is unlikely to differ from that of a bubble of fixed volume. 

We also investigated the effect of flow rate for $r_Q = 0.8$ and $r_Q = 1.2$. At low flow rates ($Q<Q_c$, see figure \ref{fig:mapexp}), time evolutions are similar to that of bubbles with $r_Q = 0.54$. For $Q>Q_c$, the main differences with bubbles of size $r_Q = 0.54$ are for time evolutions leading to break up into two bubbles of comparable sizes. For $Q \ge 0.025$, this happens for $d>d_3$ or for $d<d_3$ and $Q>Q_{ts}$ in the case of centred tip-splitting (see figure \ref{fig:mapexp}). While at $r_Q = 0.54$ the two bubbles remained on opposite sides of the rail (see figures \ref{fig:map_horizontal}(d--f) and \ref{fig:bug_break}(b)), the smaller trailing bubble often broke into two parts after elongation across the rail at $r_Q = 0.8$, leading to a state with potentially three separating bubbles. In contrast, at $r_Q = 1.2$, the smaller trailing bubble crossed the rail, leading to a single off-rail bubble. This is presumably due to the faster retraction of the rear part of the leading bubble when its size increases, which results in lower pressures at its rear, thus intensifying its attraction of the trailing bubble.


Experiments with air fingers (semi-infinite bubbles) yield similar modes of propagation to those shown in figure \ref{fig:bifurcation_exp} and the same initial deformation of the finger tip as in figure \ref{fig:mapexp}. When initial deformation leads to two or three tips along the finger front, only the fastest ultimately propagates while the other(s) stop growing and retract slowly. This simplifies the evolution because break up and aggregation events do not usually occur.

\section{Onset of oscillations}
\label{sec:Supercritical Hopf bifurcation}

\begin{figure}
	\centerline{\includegraphics[scale=1]{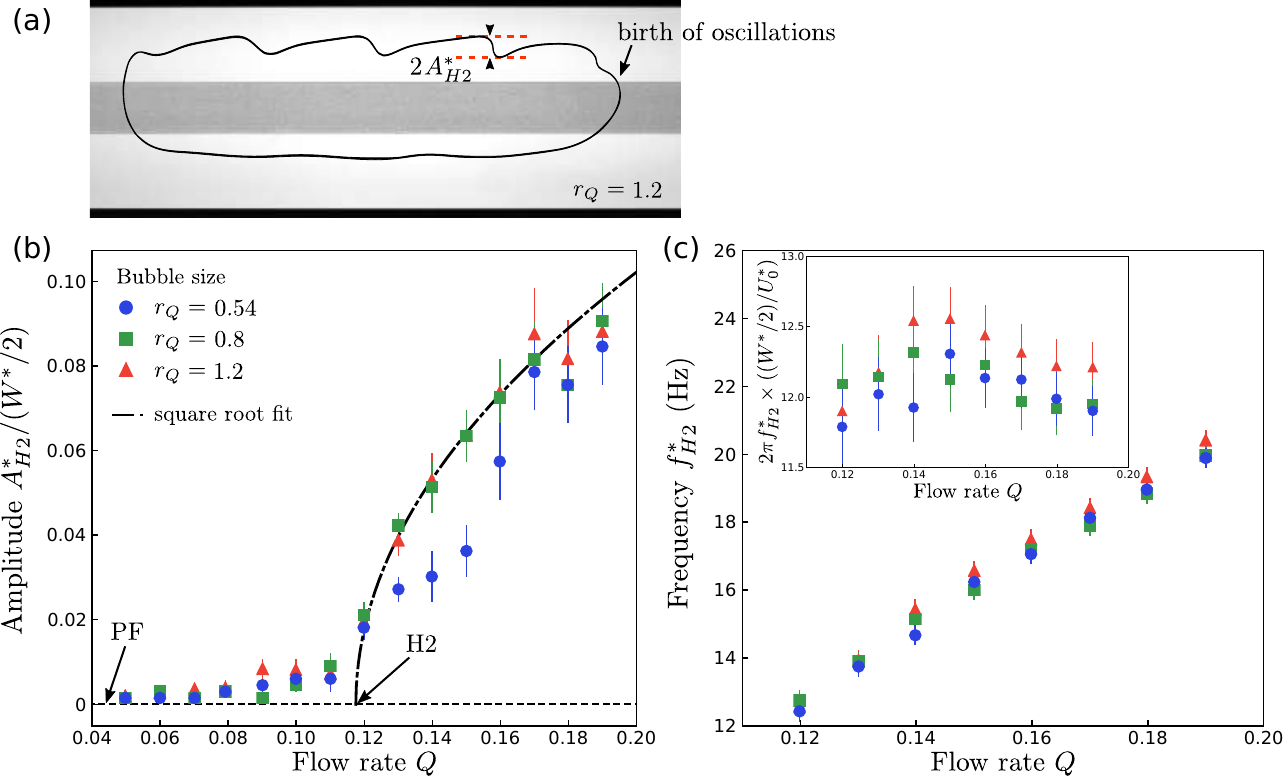}}
	\caption{(a) Snapshot of an oscillating asymmetric bubble of in-flow size $r_Q=1.2$ propagating at flow rate $Q=0.17$ (PO branch in figure \ref{fig:bifurcation_exp}). (b) Non-dimensional amplitude of oscillation $2 A_{H2}^*/W^*$ and (c) frequency $f_{H2}^*$ of oscillation against non-dimensional flow rate $Q$ for bubbles of in-flow size $r_Q = 0.54$, $0.8$ and $1.2$. A square-root fit is shown for the two largest bubbles in (a). The inset of figure (c) shows the non-dimensional frequency $2 \upi f_{H2}^* ((W^*/2)/U_0^*)$.}
	\label{fig:Hopf2}
\end{figure}

Figure \ref{fig:Hopf2}(a) shows a snapshot of an oscillating asymmetric bubble of size $r_Q=1.2$ propagating at flow rate $Q=0.17$ (PO branch in figure \ref{fig:bifurcation_exp}). The dimples periodically generated at the bubble tip (where the bubble front meets the edge of the rail) remain at fixed positions $x^*$ in the laboratory frame of reference. Each generated dimple progressively flattens under the effect of capillary pressure gradients and, consequently, the bubble is slightly flatter at the rear than at the front. We define the oscillation amplitude $A_{H2}^*$ as the maximum peak-to-peak amplitude of the interface perturbation, see figure \ref{fig:Hopf2}(a). 

The plateau value of the oscillation amplitude $2A_{H2}^*/W^*$ (reached following transient increase) is plotted in figure \ref{fig:Hopf2}(b) as a function of the non-dimensional flow rate $Q$ for three bubble sizes. The small amplitude values measured at low flow rates indicate the increasing excitability of the system near the transition which leads to damped oscillations of the bubble. The amplitude growth at higher flow rate is accurately captured by a square-root dependence on $Q$ for the two highest bubbles sizes, which yields a critical flow rate $Q_{H2} = 0.118$.

The experimental oscillation frequency $f_{H2}^*$ reported in figure \ref{fig:Hopf2}(b) increases linearly with flow rate and only slightly with bubble size. The associated non-dimensional pulsation $2 \upi f_{H2}^* ((W^*/2)/U_0^*)$ presented in the inset of figure \ref{fig:Hopf2}(b) takes an approximate average value of $12.1$. Our experimental resolution is not high enough to capture the expected pulsation increase near the bifurcation, which was calculated by \citet{keeler2019influence}.

\FloatBarrier

\bibliographystyle{jfm}
\bibliography{biblio}

\begin{thebibliography}{42}
\expandafter\ifx\csname natexlab\endcsname\relax\def\natexlab#1{#1}\fi
\def\au#1{#1} \def\ed#1{#1} \def\yr#1{#1}\def\at#1{#1}\def\jt#1{\textit{#1}}
  \def\bt#1{#1}\def\bvol#1{\textbf{#1}} \def\vol#1{#1} \def\pg#1{#1}
  \def\publ#1{#1}\def\arxiv#1{#1}\def\org#1{#1}\def\st#1{\textit{#1}}

\bibitem[Anna(2016)]{Anna2016}
{\sc \au{Anna, S.~L.}} \yr{2016}  \at{Droplets and bubbles in microfluidic
  devices}.  \jt{Annu. Rev. Fluid Mech.}  \bvol{48}~(1),  \pg{285--309}.

\bibitem[Aussillous \& Qu\'er\'e(2000)]{Aussillous2000}
{\sc \au{Aussillous, P.} \& \au{Qu\'er\'e, D.}} \yr{2000}  \at{Quick deposition
  of a fluid on the wall of a tube}.  \jt{Phys. Fluids}  \bvol{12},
  \pg{2367--2371}.

\bibitem[Bodenschatz {\em et~al.\/}(2000)Bodenschatz, Pesch \& Ahlers]{Ahlers}
{\sc \au{Bodenschatz, E.}, \au{Pesch, W.} \& \au{Ahlers, G.}} \yr{2000}
  \at{Recent developments in {R}ayleigh--{B}énard convection}.  \jt{Annu. Rev.
  Fluid Mech.}  \bvol{32}~(1),  \pg{709--778}.

\bibitem[Bretherton(1961)]{bretherton1961motion}
{\sc \au{Bretherton, F.~P.}} \yr{1961}  \at{The motion of long bubbles in
  tubes}.  \jt{J. Fluid Mech.}  \bvol{10}~(2),  \pg{166--188}.

\bibitem[Brun {\em et~al.\/}(2013)Brun, Nagel \& Gallaire]{brun2013generic}
{\sc \au{Brun, P.-T.}, \au{Nagel, M.} \& \au{Gallaire, F.}} \yr{2013}
  \at{Generic path for droplet relaxation in microfluidic channels}.  \jt{Phys.
  Rev. E}  \bvol{88}~(4),  \pg{043009}.

\bibitem[Budanur {\em et~al.\/}(2017)Budanur, Short, Farazmand, Willis \&
  Cvitanovi\'{c}]{budanur2017}
{\sc \au{Budanur, N.~B.}, \au{Short, K.~Y.}, \au{Farazmand, M.}, \au{Willis,
  A.~P.} \& \au{Cvitanovi\'{c}, P.}} \yr{2017}  \at{Relative periodic orbits
  form the backbone of turbulent pipe flow}.  \jt{J. Fluid Mech.}  \bvol{833},
  \pg{274--301}.

\bibitem[Couder(2000)]{Couder2000}
{\sc \au{Couder, Y.}} \yr{2000}  \at{Viscous fingering as an archetype for
  growth patterns.}  \bt{In {\em Perspectives in Fluid Dynamics\/}},  \pg{pp.
  53--104}.  \publ{Cambridge University Press}.

\bibitem[Cvitanovi\'{c}(2013)]{cvitanovic_2013}
{\sc \au{Cvitanovi\'{c}, P.}} \yr{2013}  \at{Recurrent flows: the clockwork
  behind turbulence}.  \jt{J. Fluid Mech.}  \bvol{726},  \pg{1--4}.

\bibitem[Duguet {\em et~al.\/}(2008)Duguet, Willis \& Kerswell]{duguet2008}
{\sc \au{Duguet, Y.}, \au{Willis, A.~P.} \& \au{Kerswell, R.~R.}} \yr{2008}
  \at{Transition in pipe flow: the saddle structure on the boundary of
  turbulence}.  \jt{J. Fluid Mech.}  \bvol{613},  \pg{255--274}.

\bibitem[Eckhardt {\em et~al.\/}(2008)Eckhardt, Faisst, Schmiegel \&
  Schneider]{eckhardt2008dynamical}
{\sc \au{Eckhardt, B.}, \au{Faisst, H.}, \au{Schmiegel, A.} \& \au{Schneider,
  T.~M.}} \yr{2008}  \at{Dynamical systems and the transition to turbulence in
  linearly stable shear flows}.  \jt{Phil. Trans. R. Soc. Lond. A}
  \bvol{366}~(1868),  \pg{1297--1315}.

\bibitem[Farano {\em et~al.\/}(2019)Farano, Cherubini, Robinet, De~Palma \&
  Schneider]{farano_etal_2019}
{\sc \au{Farano, M.}, \au{Cherubini, S.}, \au{Robinet, J.-C.}, \au{De~Palma,
  P.} \& \au{Schneider, T.~M.}} \yr{2019}  \at{Computing heteroclinic orbits
  using adjoint-based methods}.  \jt{J. Fluid Mech.}  \bvol{858},  \pg{R3}.

\bibitem[Franco-G{\'o}mez {\em et~al.\/}(2016)Franco-G{\'o}mez, Thompson, Hazel
  \& Juel]{franco2016sensitivity}
{\sc \au{Franco-G{\'o}mez, A.}, \au{Thompson, A.~B.}, \au{Hazel, A.~L.} \&
  \au{Juel, A.}} \yr{2016}  \at{Sensitivity of {S}affman--{T}aylor fingers to
  channel-depth perturbations}.  \jt{J.~Fluid Mech.}  \bvol{794},
  \pg{343--368}.

\bibitem[Franco-G{\'o}mez {\em et~al.\/}(2017)Franco-G{\'o}mez, Thompson, Hazel
  \& Juel]{franco2017bubble}
{\sc \au{Franco-G{\'o}mez, A.}, \au{Thompson, A.~B.}, \au{Hazel, A.~L.} \&
  \au{Juel, A.}} \yr{2017}  \at{Bubble propagation on a rail: a concept for
  sorting bubbles by size}.  \jt{Soft Matter}  \bvol{13}~(46),
  \pg{8684--8697}.

\bibitem[Franco-G{\'o}mez {\em et~al.\/}(2018)Franco-G{\'o}mez, Thompson, Hazel
  \& Juel]{franco2018bubble}
{\sc \au{Franco-G{\'o}mez, A.}, \au{Thompson, A.~B.}, \au{Hazel, A.~L.} \&
  \au{Juel, A.}} \yr{2018}  \at{Bubble propagation in {H}ele-{S}haw channels
  with centred constrictions}.  \jt{Fluid Dyn. Res.}  \bvol{50}~(2),
  \pg{021403}.

\bibitem[Gallino {\em et~al.\/}(2018)Gallino, Schneider \&
  Gallaire]{gallino2018edge}
{\sc \au{Gallino, G.}, \au{Schneider, T.~M.} \& \au{Gallaire, F.}} \yr{2018}
  \at{Edge states control droplet breakup in subcritical extensional flows}.
  \jt{Phys. Rev. Fluid}  \bvol{3},  \pg{073603}.

\bibitem[Gelfgat(2019)]{gelfgat_bif}
{\sc \au{Gelfgat, A.}}, ed. \yr{2019} {\em Computational Modelling of
  Bifurcations and Instabilities in Fluid Dynamics\/}.  \publ{Springer}.

\bibitem[Gollub(1995)]{Gollub1995}
{\sc \au{Gollub, J.~P.}} \yr{1995}  \at{Order and disorder in fluid motion}.
  \jt{Proc. Natl. Acad. Sci. U.S.A.}  \bvol{92}~(15),  \pg{6705--6711}.

\bibitem[Gollub \& Swinney(1975)]{Swinney}
{\sc \au{Gollub, J.~P.} \& \au{Swinney, H.~L.}} \yr{1975}  \at{Onset of
  turbulence in a rotating fluid}.  \jt{Phys. Rev. Lett.}  \bvol{35},
  \pg{927--930}.

\bibitem[Heil \& Hazel(2006)]{heil2006oomph}
{\sc \au{Heil, M.} \& \au{Hazel, A.~L.}} \yr{2006}  \at{oomph-lib--an
  object-oriented multi-physics finite-element library}.  \bt{In {\em
  Fluid-structure interaction\/}},  \pg{pp. 19--49}.  \publ{Springer}.

\bibitem[Hopf(1948)]{Hopf1948}
{\sc \au{Hopf, E.}} \yr{1948}  \at{A mathematical example displaying features
  of turbulence}.  \jt{Commun. Pur. Appl. Math.}  \bvol{1}~(4),  \pg{303--322}.

\bibitem[Jisiou {\em et~al.\/}(2014)Jisiou, Dawson, Thompson, Mohr, Fielden,
  Hazel \& Juel]{jisiou2014geometry}
{\sc \au{Jisiou, M.}, \au{Dawson, G.}, \au{Thompson, A.~B.}, \au{Mohr, S.},
  \au{Fielden, P.~R.}, \au{Hazel, A.~L.} \& \au{Juel, A.}} \yr{2014}
  \at{Geometry-induced oscillations of finite bubbles in microchannels}.
  \jt{Procedia IUTAM}  \bvol{11},  \pg{81--88}.

\bibitem[Kawahara {\em et~al.\/}(2012)Kawahara, Uhlmann \& van{
  }Veen]{kawahara2012}
{\sc \au{Kawahara, G.}, \au{Uhlmann, M.} \& \au{van{ }Veen, L.}} \yr{2012}
  \at{The significance of simple invariant solutions in turbulent flows}.
  \jt{Annu. Rev. Fluid Mech.}  \bvol{44}~(1),  \pg{203--225}.

\bibitem[Keeler {\em et~al.\/}(2020)Keeler, Thompson, Gaillard, Juel \&
  Hazel]{Keeler2020}
{\sc \au{Keeler, J.~S.}, \au{Thompson, A.~B.}, \au{Gaillard, A.}, \au{Juel, A.}
  \& \au{Hazel, A.~L.}} \yr{2020} The interaction of multiple bubbles in a
  {H}ele-{S}haw channel. In preparation.

\bibitem[Keeler {\em et~al.\/}(2019)Keeler, Thompson, Lemoult, Juel \&
  Hazel]{keeler2019influence}
{\sc \au{Keeler, J.~S.}, \au{Thompson, A.~B.}, \au{Lemoult, G.}, \au{Juel, A.}
  \& \au{Hazel, A.~L.}} \yr{2019}  \at{The influence of invariant solutions on
  the transient behaviour of an air bubble in a {H}ele-{S}haw channel}.
  \jt{Proc. R. Soc. A}  \bvol{475}~(2232),  \pg{20190434}.

\bibitem[Kerswell(2005)]{kerswell2005recent}
{\sc \au{Kerswell, R.~R.}} \yr{2005}  \at{Recent progress in understanding the
  transition to turbulence in a pipe}.  \jt{Nonlinearity}  \bvol{18}~(6),
  \pg{R17}.

\bibitem[Kopf‐Sill \& Homsy(1988)]{homsy}
{\sc \au{Kopf‐Sill, A.~R.} \& \au{Homsy, G.~M.}} \yr{1988}  \at{Bubble motion
  in a {H}ele–{S}haw cell}.  \jt{Phys. Fluids}  \bvol{31}~(1),  \pg{18--26}.

\bibitem[Lorenz(1963)]{Lorenz1963}
{\sc \au{Lorenz, E.~N.}} \yr{1963}  \at{Deterministic nonperiodic flow}.
  \jt{J. Atmos. Sci.}  \bvol{20}~(2),  \pg{130--141}.

\bibitem[Manneville(1990)]{Manneville}
{\sc \au{Manneville, P.}} \yr{1990} {\em Dissipative Structures and Weak
  Turbulence\/}.  \publ{Boston: Academic Press}.

\bibitem[McLean \& Saffman(1981)]{mccleantension}
{\sc \au{McLean, J.~W.} \& \au{Saffman, P.~G.}} \yr{1981}  \at{The effect of
  surface tension on the shape of fingers in a {H}ele-{S}haw cell}.  \jt{J.
  Fluid Mech.}  \bvol{102},  \pg{455--469}.

\bibitem[Mullin(1993)]{Mullin}
{\sc \au{Mullin, T.}}, ed. \yr{1993} {\em The nature of chaos\/}.  \publ{OUP}.

\bibitem[Pailha {\em et~al.\/}(2012)Pailha, Hazel, Glendinning \&
  Juel]{pailha2012oscillatory}
{\sc \au{Pailha, M.}, \au{Hazel, A.~L.}, \au{Glendinning, P.~A.} \& \au{Juel,
  A.}} \yr{2012}  \at{Oscillatory bubbles induced by geometrical constraint}.
  \jt{Phys. Fluids}  \bvol{24}~(2),  \pg{021702}.

\bibitem[Park \& Homsy(1984)]{park1984two}
{\sc \au{Park, C.-W.} \& \au{Homsy, G.~M.}} \yr{1984}  \at{Two-phase
  displacement in {H}ele-{S}haw cells: theory}.  \jt{J. Fluid Mech.}
  \bvol{139},  \pg{291--308}.

\bibitem[Ruelle \& Takens(1971)]{Ruelle}
{\sc \au{Ruelle, D.} \& \au{Takens, F.}} \yr{1971}  \at{On the nature of
  turbulence}.  \jt{Commun. Math. Phys.}  \bvol{20},  \pg{167--192}.

\bibitem[Saffman \& Taylor(1958)]{saffman1958penetration}
{\sc \au{Saffman, P.~G.} \& \au{Taylor, G.~I.}} \yr{1958}  \at{The penetration
  of a fluid into a porous medium or {H}ele-{S}haw cell containing a more
  viscous liquid}.  \jt{Phil. Trans. R. Soc. Lond.}  \bvol{245}~(1242),
  \pg{312--329}.

\bibitem[S\'anchez~Umbr\'{\i}a \& Net(2019)]{Sanchez_2019}
{\sc \au{S\'anchez~Umbr\'{\i}a, J.} \& \au{Net, M.}} \yr{2019}  \at{Torsional
  solutions of convection in rotating fluid spheres}.  \jt{Phys. Rev. Fluid}
  \bvol{4},  \pg{013501}.

\bibitem[Schneider {\em et~al.\/}(2007)Schneider, Eckhardt \&
  Yorke]{schneider2007turbulence}
{\sc \au{Schneider, T.~M.}, \au{Eckhardt, B.} \& \au{Yorke, J.~A.}} \yr{2007}
  \at{Turbulence transition and the edge of chaos in pipe flow}.  \jt{Phys.
  Rev. Lett.}  \bvol{99}~(3),  \pg{034502}.

\bibitem[Skufca {\em et~al.\/}(2006)Skufca, Yorke \& Eckhardt]{Edge_track}
{\sc \au{Skufca, J.~D.}, \au{Yorke, J.~A.} \& \au{Eckhardt, B.}} \yr{2006}
  \at{Edge of chaos in a parallel shear flow}.  \jt{Phys. Rev. Lett.}
  \bvol{96},  \pg{174101}.

\bibitem[Tabeling {\em et~al.\/}(1987)Tabeling, Zocchi \&
  Libchaber]{tabeling1987experimental}
{\sc \au{Tabeling, P.}, \au{Zocchi, G.} \& \au{Libchaber, A.}} \yr{1987}
  \at{An experimental study of the {S}affman-{T}aylor instability}.  \jt{J.
  Fluid Mech.}  \bvol{177},  \pg{67--82}.

\bibitem[Tanveer \& Saffman(1987)]{tanveer1987stability}
{\sc \au{Tanveer, S.} \& \au{Saffman, P.~G.}} \yr{1987}  \at{Stability of
  bubbles in a {H}ele--{S}haw cell}.  \jt{Phys. Fluids}  \bvol{30}~(9),
  \pg{2624--2635}.

\bibitem[Thompson {\em et~al.\/}(2014)Thompson, Juel \&
  Hazel]{thompson2014multiple}
{\sc \au{Thompson, A.~B.}, \au{Juel, A.} \& \au{Hazel, A.~L.}} \yr{2014}
  \at{Multiple finger propagation modes in {H}ele-{S}haw channels of variable
  depth}.  \jt{J.~Fluid Mech.}  \bvol{746},  \pg{123--164}.

\bibitem[Tuckerman {\em et~al.\/}(2019)Tuckerman, Langham \&
  Willis]{Tuckerman2019}
{\sc \au{Tuckerman, L.~S.}, \au{Langham, J.} \& \au{Willis, A.}} \yr{2019}
  \at{Order-of-magnitude speedup for steady states and traveling waves via
  {S}tokes preconditioning in channelflow and openpipeflow}.  \bt{In {\em
  Computational Modelling of Bifurcations and Instabilities in Fluid
  Dynamics\/} (ed. \ed{A.~Gelfgat})},  \pg{pp. 3--31}.  \publ{Springer
  International Publishing}.

\bibitem[Vaquero-Stainer {\em et~al.\/}(2019)Vaquero-Stainer, Heil, Juel \&
  Pihler-Puzovi\ifmmode~\acute{c}\else \'{c}\fi{}]{Christian}
{\sc \au{Vaquero-Stainer, C.}, \au{Heil, M.}, \au{Juel, A.} \&
  \au{Pihler-Puzovi\ifmmode~\acute{c}\else \'{c}\fi{}, D.}} \yr{2019}
  \at{Self-similar and disordered front propagation in a radial {H}ele-{S}haw
  channel with time-varying cell depth}.  \jt{Phys. Rev. Fluid}  \bvol{4},
  \pg{064002}.

\end{thebibliography}

\end{document}